\documentclass[aps,prd,twocolumn,preprintnumbers,superscriptaddress,nofootinbib,floatfix]{revtex4-1}

\usepackage{amssymb,amsmath,amsfonts}
\usepackage{graphicx}
\usepackage{epstopdf}
\usepackage{hyperref}
\usepackage{color}
\definecolor{steelblue}{RGB}{25,25,112}
\definecolor{DarkGreen}{rgb}{0.0,0.5,0.0}
\hypersetup{colorlinks,linkcolor={blue},citecolor={blue},urlcolor={steelblue}}
\DeclareRobustCommand{\okina}{%
  \raisebox{\dimexpr\fontcharht\font`A-\height}{%
    \scalebox{0.8}{`}%
  }%
}

\newcommand{\mpl}{M_{\rm Pl}}

\newcommand{\dd}{\mathrm{d}}

\begin{document}

\title{Direct detection of dark energy: the XENON1T excess and future prospects}

\author{Sunny Vagnozzi}
\email{sunny.vagnozzi@ast.cam.ac.uk}
\affiliation{Kavli Institute for Cosmology (KICC), University of Cambridge, Madingley Road, Cambridge CB3 0HA, United Kingdom}
\thanks{S.V. and L.V. contributed equally to this work}
\affiliation{Institute of Astronomy (IoA), University of Cambridge, Madingley Road, Cambridge CB3 0HA, United Kingdom}

\author{Luca Visinelli}
\email{luca.visinelli@sjtu.edu.cn}
\affiliation{Istituto Nazionale di Fisica Nucleare (INFN), Laboratori Nazionali di Frascati, C.P. 13, I-100044 Frascati, Italy}
\thanks{S.V. and L.V. contributed equally to this work}
\affiliation{Tsung-Dao Lee Institute (TDLI), Shanghai Jiao Tong University, 200240 Shanghai, China}
\affiliation{Gravitation Astroparticle Physics Amsterdam (GRAPPA), University of Amsterdam, Science Park 904, 1098 XH Amsterdam, The Netherlands}

\author{Philippe Brax}
\email{philippe.brax@cea.fr}
\affiliation{Institute de Physique The{\'o}rique (IPhT), Universit{\'e} Paris-Saclay, CNRS, CEA, F-91191, Gif-sur-Yvette Cedex, France}

\author{Anne-Christine Davis}
\email{ad107@cam.ac.uk}
\affiliation{Department of Applied Mathematics and Theoretical Physics (DAMTP), Center for Mathematical Sciences, University of Cambridge, CB3 0WA, United Kingdom}
\affiliation{Kavli Institute for Cosmology (KICC), University of Cambridge, Madingley Road, Cambridge CB3 0HA, United Kingdom}

\author{Jeremy Sakstein}
\email{sakstein@hawaii.edu}
\affiliation{Department of Physics \& Astronomy, University of Hawai\okina i, Watanabe Hall, 2505 Correa Road, Honolulu, HI, 96822, USA}

\date{\today}
 
\begin{abstract}
\noindent We explore the prospects for direct detection of dark energy by current and upcoming terrestrial dark matter direct detection experiments. If dark energy is driven by a new light degree of freedom coupled to matter and photons then dark energy quanta are predicted to be produced in the Sun. These quanta free-stream towards Earth where they can interact with Standard Model particles in the detection chambers of direct detection experiments, presenting the possibility that these experiments could be used to test dark energy. Screening mechanisms, which suppress fifth forces associated with new light particles, and are a necessary feature of many dark energy models, prevent production processes from occurring in the core of the Sun, and similarly, in the cores of red giant, horizontal branch, and white dwarf stars. Instead, the coupling of dark energy to photons leads to production in the strong magnetic field of the solar tachocline via a mechanism analogous to the Primakoff process. This then allows for detectable signals on Earth while evading the strong constraints that would typically result from stellar probes of new light particles. As an example, we examine whether the electron recoil excess recently reported by the XENON1T collaboration can be explained by chameleon-screened dark energy, and find that such a model is preferred over the background-only hypothesis at the $2.0\sigma$ level, in a large range of parameter space not excluded by stellar (or other) probes. This raises the tantalizing possibility that XENON1T may have achieved the first direct detection of dark energy. Finally, we study the prospects for confirming this scenario using planned future detectors such as XENONnT, PandaX-4T, and LUX-ZEPLIN.
\end{abstract}

\maketitle

\section{Introduction}
\label{sec:Intro}

More than two decades after the discovery that the expansion of the Universe is accelerating~\cite{Riess:1998cb,Perlmutter:1998np}, the nature of the \textit{dark energy} (DE) component driving this phenomenon and making up $\sim 70\%$ of the energy budget of the Universe remains a mystery~\cite{Nojiri:2006ri,Frieman:2008sn,Bamba:2012cp,Joyce:2014kja,Huterer:2017buf}. What is perhaps the simplest theoretical DE candidate, a cosmological constant (CC) resulting from the collective zero-point energy of quantum fields, suffers from a severe disagreement between its theoretical value suggested from quantum field theory considerations, and the tiny value inferred from cosmological observations~\cite{Sola:2013gha,Burgess:2013ara,Padilla:2015aaa}. This staggering discrepancy goes under the name of \textit{cosmological constant problem}~\cite{Burgess:2013ara, Padilla:2015aaa}. Several DE models have been proposed as alternatives to the CC, within most of which the cosmic acceleration phenomenon is ascribed to additional (typically scalar) degrees of freedom, either in the form of new fundamental particles and fields, or modifications to General Relativity (GR). Competing models predict a variety of novel cosmological phenomena which will be actively sought for by the next generations of cosmological surveys, including but not limited to the Simons Observatory~\cite{Ade:2018sbj,Abitbol:2019nhf}, CMB-S4~\cite{Abazajian:2016yjj,Abazajian:2019tiv}, Euclid~\cite{Laureijs:2011gra,Amendola:2016saw}, DESI~\cite{Aghamousa:2016zmz}, the Vera C.\ Rubin Observatory Legacy Survey of Space and Time~\cite{Abate:2012za}, and the Nancy Grace Roman Space Telescope~\cite{Spergel:2013tha}. 

In all but the simplest models, the scalar field leads to a modification of gravity where the associated new light scalar degree of freedom couples to matter~\cite{Sotiriou:2008rp,Clifton:2011jh,Nojiri:2017ncd}. Such theories predict the existence of \textit{fifth forces} that are ostensibly excluded by solar system tests of GR~\cite{Will:2005va, Sakstein:2017pqi}. To circumvent this problem, realistic models inevitably need to include some form of \textit{screening mechanism}~\cite{Khoury:2010xi,Brax:2013ida,Joyce:2014kja, Sakstein:2018fwz, Baker:2019gxo}. Screening mechanisms dynamically suppress fifth forces in the solar system without the need to tune the model parameters, resorting to environmental effects in the case of several (but not all) of these mechanisms. This then allows for strong deviations from GR on cosmological scales, which could modify the growth of structure. The ability to  account simultaneously for cosmic acceleration and satisfy solar system tests of GR have made modified gravity (MG) theories with screening mechanisms leading science targets for current and upcoming cosmological surveys~\cite{Burrage:2017qrf}. In addition to their indirect effects on the cosmological background expansion and structure formation~\cite{Koyama:2015vza, Ferreira:2019xrr},\footnote{See e.g.\ Refs.~\cite{Sakstein:2019qgn,Vagnozzi:2019kvw, Jimenez:2020ysu,Berghaus:2020ekh,Cai:2021wgv} for recent works studying the effect of direct couplings of DE to baryons on both local and cosmological observables.} DE and MG theories with screening mechanisms have the attractive feature that they are also amenable to direct detection of their associated effects. The distinctive fifth forces they predict can be searched for in laboratory experiments~\cite{Burrage:2016bwy, Burrage:2017qrf,Homma:2019rqb} and astrophysical environments~\cite{Chang:2010xh,Iorio:2011ay,Llinares:2012ds,Llinares:2013jua,Gronke:2014gaa,Gronke:2015ama,Santos:2016rdg,Katsuragawa:2016yir,Katsuragawa:2017wge,Sakstein:2018fwz,Olmo:2019flu,Baker:2019gxo,KumarPoddar:2020kdz,Straight:2020zke,Tsai:2021irw}, and their observation would unequivocally confirm the hypothesis that DE is linked to a modification of GR. 

The purpose of this paper is to broaden the scope of new physics accessible to terrestrial dark matter (DM) direct detection experiments by exploring their potential to detect DE quanta produced in the Sun, and to open up a new frontier for the direct detection of dark energy. Screening mechanisms suppress the production of DE quanta in the core of the Sun due to the large scalar mass or weak coupling to matter, but they can be produced in regions of strongly magnetised plasmas through couplings of the scalar to photons. In particular, they can be produced in the solar tachocline,\footnote{The tachocline is the turbulent shear layer located at the base of the solar convection zone, marking the transition between the radiative interior and the differentially rotating convective zone~\cite{1992A&A...265..106S}. The radial position of the tachocline is approximately $0.7R_{\odot}$, with $R_{\odot}$ the solar radius.} in a manner analogous to the Primakoff process for axions~\cite{Brax:2010xq,Brax:2011wp}. This possibility opens up yet another exciting avenue for directly detecting DE fluctuations:  DE particles produced in the solar tachocline could be observed in dark matter (DM) direct detection experiments. To date, searches for DE-like particles in direct detection experiments have mostly focused on axion-specific experiments such as the CERN Axion Solar Telescope (CAST)~\cite{Baum:2014rka,Anastassopoulos:2015yda,Anastassopoulos:2018kcs,Cuendis:2019mgz} and the Axion Dark Matter Experiment (ADMX)~\cite{Rybka:2010ah}. Other detection techniques currently being considered are at different stages of realisation, or have started to gather data~\cite{Vogel:2013bta, Kahn:2016aff, TheMADMAXWorkingGroup:2016hpc, Barbieri:2016vwg, Brubaker:2016ktl, Alesini:2017ifp, Alesini:2019nzq}. For additional details, see Refs.~\cite{Irastorza:2018dyq, Visinelli:2018utg, DiLuzio:2020wdo, Sikivie:2020zpn}.

All well-known screening mechanisms can be classified into chameleon~\cite{Khoury:2003rn}, symmetron~\cite{Hinterbichler:2010es}, Damour-Polyakov~\cite{Damour:1994zq}, K-mouflage~\cite{Babichev:2009ee}, and Vainshtein~\cite{Vainshtein:1972sx} screened theories. In order to encapsulate the diversity exhibited by screening mechanisms, we will use an effective theory for the fluctuations of the scalar about the background profile {due to the environment} that includes operators relevant for each mechanism, including generic couplings to the standard model (SM). Using this, we will lay out the formalism for computing the expected signal in the associated detectors following from the scattering/absorption of DE scalars produced in the solar tachocline by nuclei or electrons (depending on the specific detector details). We will assume that the DE scalar possesses a coupling to photons, allowing for its production from photons in the tachocline. We will also use the environmental dependence of the couplings in the effective theory for the scalar fluctuations. We note that screening mechanisms of the chameleon, symmetron and Damour-Polyakov types show a clear dependence of these coupling on the local density, i.e.\ on the environment. On the other hand, screening mechanisms of the K-mouflage and Vainshtein types depend on more global features of the matter distribution, such as the total mass of the stars, with little or no dependence on the local matter density~\cite{Koyama:2015oma}. In order to provide a well-studied and well-constrained example, we will later specialise to the chameleon mechanism. Theories that screen using the chameleon mechanism~\cite{Khoury:2003aq,Khoury:2003rn,Brax:2004qh,Burrage:2017qrf} do so by increasing the mass of the DE scalar in dense environments, making their fifth force too short-ranged to be relevant in the solar system or in terrestrial fifth force searches.

As a case study, we will apply our methodology to the XENON1T DM direct detection experiment~\cite{Aprile:2017aty}, which recently reported a $\sim 3.3\sigma$ excess in the electron recoil data above their expected background, in the energy range $1-7\,$keV~\cite{Aprile:2020tmw}. The XENON1T collaboration finds that a fit to the signal which includes a solar axion component is preferred over the background-only hypothesis at a significance of $3.4\sigma$. Alternative explanations proposed by the collaboration include an additional tritium background and a neutrino magnetic moment, both of which are preferred with a significance of $3.2\sigma$. Unfortunately, for the solar axion (and, to a minor but still important extent, for the neutrino magnetic moment), the parameters that best fit the XENON1T signal are strongly excluded by astrophysical observations of stellar evolution~\cite{DiLuzio:2020jjp}, particularly by constraints from the cooling of horizontal branch stars in globular clusters, the cooling of white dwarfs, and the tip of the red giant branch $I$-band magnitude of globular clusters and galaxies. The difficulties faced by these three scenarios have spurred the proposal of various new physics models that might also be able to explain the XENON1T excess, with varying degree of motivation or plausibility: for a list of works in this direction, we refer the reader to e.g.\ Refs.~\cite{DiLuzio:2020jjp,Takahashi:2020bpq,Kannike:2020agf,Alonso-Alvarez:2020cdv,Fornal:2020npv,Boehm:2020ltd,Harigaya:2020ckz,Bally:2020yid,Su:2020zny,Du:2020ybt,Chen:2020gcl,Dey:2020sai,Bell:2020bes,Buch:2020mrg,AristizabalSierra:2020edu,Choi:2020udy,Paz:2020pbc,Lee:2020wmh,Cao:2020bwd,Khan:2020vaf,Nakayama:2020ikz,Bramante:2020zos,An:2020bxd,Gao:2020wer,Lindner:2020kko,Bloch:2020uzh,McKeen:2020vpf,Ge:2020jfn,Chao:2020yro,Baek:2020owl,Chigusa:2020bgq,Miranda:2020kwy,Okada:2020evk,Choi:2020kch,He:2020wjs,Shoemaker:2020kji,Long:2020uyf,Chiang:2020hgb,Arcadi:2020zni,Ema:2020fit,Kim:2020aua,Cao:2020oxq,Borah:2020jzi,Farzan:2020dds,Zu:2020bsx,Karozas:2020pun,Chakraborty:2020vec,Keung:2020uew,Foot:2020ehn,Adams:2020ejt,Aboubrahim:2020iwb,Buttazzo:2020vfs,He:2020sat,Xu:2020qsy,Dutta:2021wbn}.

Applying our formalism to XENON1T, we find that chameleon-screened DE provides a fit to the signal of only slightly lower quality than the aforementioned scenarios, and is preferred over the background-only hypothesis at a significance of $2.0\sigma$. The stellar bounds that are debilitating for the solar axion interpretation of the XENON1T signal are avoided by chameleon-screened DE theories due to the associated fifth force screening, as well as by the dependence of the mass of the scalar field on the local density of matter~\cite{Brax:2010xq, Brax:2011wp}. In particular, the astrophysical objects from which the strongest axion bounds derive --- red giant, horizontal branch, and white dwarf stars --- have cores whose density is larger than that of the solar tachocline by a factor of $>10^6$. As a consequence, chameleon-screened DE scalars are too heavy to be thermally produced within these objects, and cannot act as a new source of energy loss. This raises the tantalizing possibility that XENON1T, originally devised to detect DM, may instead have detected dark energy, or in any case a propagating scalar degree of freedom of a theory of modified gravity.

The possibility that chameleon-screened scalars might indeed be at the origin of the XENON1T signal is independently testable by a number of upcoming low-threshold DM direct detection experiments. These experiments are all sensitive to recoil energies of ${\cal O}(100\,{\rm eV})$ or lower, and are expected to detect the chameleon-induced signal at much higher statistical significance than XENON1T. Examples of experiments in this class, some of which make use of cryogenic semiconductor detectors, include SuperCDMS~\cite{Agnese:2016cpb}, CDEX-10~\cite{Jiang:2018pic}, DAMIC~\cite{Chavarria:2014ika}, CRESST-III~\cite{Abdelhameed:2019hmk}, PICO~\cite{Amole:2019fdf}, LUX-ZEPLIN~\cite{Akerib:2018lyp}, EDELWEISS~\cite{Arnaud:2020svb}, SENSEI~\cite{Abramoff:2019dfb}, PandaX-II~\cite{Wang:2020coa}, PandaX-4T~\cite{Zhang:2018xdp}, and of course XENONnT~\cite{Aprile:2020vtw} among others. We will demonstrate that these experimental setups, often discussed in the context of searches for light (sub-GeV or even MeV) DM scattering off nuclei or electrons, are equally well-placed to detect the signal of solar chameleons.\footnote{In the following we will use the word chameleon for all chameleon-screened scalars. When the original chameleon scalar with an inverse power potential is meant, this will be explicitly specified.}

The rest of this paper is organized as follows. Our effective theory for screened dark energy in the solar system is presented and discussed in Sec.~\ref{sec:chameleons}. The specific chameleon model we specialise to in this work is presented in Sec.~\ref{sec:model}, where we calculate its flux from production in the solar tachocline, and detection cross-section in DM direct detection experiments. In Sec.~\ref{sec:analysis} we discuss the data analysis method we use to confront our model against the XENON1T signal. In Sec.~\ref{sec:results} we present the results following from this analysis. We critically discuss these results and prospects for probing chameleons in current and future DM direct detection experiments in Sec.~\ref{sec:discussion}. Finally, we draw our concluding remarks in Sec.~\ref{sec:conclusions}. In addition, our paper contains three technical appendices. Appendix~\ref{sec:AppendixA} is devoted to a more detailed discussion of the chameleon mechanism. Appendix~\ref{sec:production} revises the production of chameleon-screened DE scalars in the Sun, as well as the resulting flux and energy spectrum of solar chameleons on Earth. Appendix~\ref{sec:detectioncrosssection} presents details concerning the computation of the cross-section for what we refer to as the ``chameleo-electric effect'', a process which is analogous to the photoelectric and axio-electric effects for photons and axions respectively, and which is relevant for the computation of the expected signal in the XENON1T detector. All the codes associated with this work are made publicly available online at \href{https://github.com/lucavisinelli/XENONCHAM}{github.com/lucavisinelli/XENONCHAM}.

\section{Dark Energy Effective Theory in the Solar System}
\label{sec:chameleons}

Here, we work under the assumption that dark energy arises as a result of the cosmological dynamics of a single scalar field, which we denote by $\varphi$.
This scalar does not have to be a fundamental field. Instead, it could arise as a low energy degree of freedom emerging from more involved dynamics at higher energy. For example, the scalar could be the St\"{u}ckelberg field of the broken time-diffeomorphism symmetry of the cosmological background~\cite{Gleyzes:2015pma}. Our further assumption is that this scalar is involved in generating the acceleration of the Universe and at the same time couples to gravity and/or the SM in a manner that is ghost- and pathology-free, as explicitly realised e.g.\ in models of the Horndeski~\cite{Horndeski:1974wa}, beyond Horndeski (GLPV)~\cite{Zumalacarregui:2013pma, Gleyzes:2014dya}, or Degenerate Higher-Order Scalar-Tensor (DHOST) classes of scalar-tensor theories~\cite{Langlois:2015cwa,Langlois:2018dxi}. These theories are among the leading candidate DE theories accompanying a modification of gravity (see Refs.~\cite{Nojiri:2006ri,Sotiriou:2008rp,Clifton:2011jh,Joyce:2014kja,Sebastiani:2016ras,Nojiri:2017ncd,Ferreira:2019xrr} for more general reviews concerning MG theories and cosmological tests thereof).

Multi-messenger astronomy involving a gravitational wave (GW) and an associated optical counterpart can be used to constrain the properties of theories of modified gravity. To date, there has been one such confirmed event. In 2017, the LIGO/Virgo collaboration observed the gravitational wave event GW170817~\cite{TheLIGOScientific:2017qsa}, resulting from a binary neutron star merger. Simultaneously, the \textit{Fermi} Gamma-ray Burst Monitor and the \textit{INTEGRAL} Anti-Coincidence Shield spectrometer detected the short gamma-ray burst GRB170817A~\cite{Monitor:2017mdv, Goldstein:2017mmi}, which was identified as being the electromagnetic counterpart to GW170817. The joint GW170817/GRB170817A detection restricts the speed of GWs to differ from the speed of light by no more than one part in $10^{15}$, setting strong constraints on theories of the Horndeski~\cite{Sakstein:2017xjx, Baker:2017hug, Ezquiaga:2017ekz, Creminelli:2017sry, Arai:2017hxj,Kreisch:2017uet, Kase:2018aps}, beyond Horndeski~\cite{Kase:2018iwp, Amendola:2017orw}, and DHOST types~\cite{Bartolo:2017ibw, Casalino:2018tcd,Ganz:2018vzg, Casalino:2018wnc, Arai:2019zul}, although these constraints are subject to important caveats~\cite{deRham:2018red, Copeland:2018yuh}. Other important constraints on such theories arise from potential instabilities in GW backgrounds~\cite{Creminelli:2018xsv, Creminelli:2019kjy}, astrophysical bounds~\cite{Sakstein:2015zoa, Sakstein:2015aac, Sakstein:2018fwz, Baker:2019gxo, Saltas:2019ius}, and cosmological constraints~\cite{Renk:2017rzu, Dutta:2017fjw, Peirone:2019yjs, Zumalacarregui:2020cjh}. Nevertheless, large regions of parameter space remain observationally and theoretically viable~\cite{Dima:2017pwp, Langlois:2017dyl, Crisostomi:2017lbg, Crisostomi:2017pjs, Crisostomi:2019yfo}. Since we will work at the level of the effective theory in the solar system, our formalism is insensitive to the details of the fundamental theory responsible for driving cosmic acceleration. We will thus assume that the aforementioned bounds are satisfied. The chameleon-screened dark energy theories that we study in this paper predict that the speed of gravitational waves is luminal so the bounds are automatically satisfied without the need to tune parameters. Attempts to embed our more general solar system effective theory into a  covariant model which can be extended to cosmological scales should ensure that these bounds are satisfied. We defer this study to future work. 

We begin by expanding the field $\varphi(\vec{x},t)$ around some background value $\varphi_0(\vec{x},t)$ as $\varphi(\vec{x},t)=\varphi_0(\vec{x},t) + \phi(\vec{x},t)$, where $\phi(\vec{x},t)$ is a spacetime-dependent perturbation. Depending on the nature of the fundamental theory, $\varphi_0(\vec{x},t)$ could either be the field in the cosmological background, the field sourced by the Milky Way, or even different in the Earth and the Sun. The latter scenario is realised by the chameleon screening mechanism. The relevant part of the action for studying solar phenomenology comprises two terms. The first  is the quadratic part of a scalar field action with background-dependent kinetic and mass terms
\begin{equation}
    S=\int\dd^4 x\sqrt{-g}\left[
    -\frac12Z^{\mu\nu}(\varphi_0)\partial_\mu\phi\partial_\nu\phi-
    \frac12m^2(\varphi_0)\phi^2\right ]\,,
\end{equation}
where $Z^{\mu\nu}$ is the kinetic matrix. We have not included self-interaction terms in the form of a polynomial in $\phi^n, \ n\ge 3$ as they are not relevant to the scalar production mechanism we will consider. In principle, each of the monomials in this expansion have $\varphi_0$-dependent couplings.
The second relevant part of the action is the coupling to matter and photons
\begin{eqnarray}
    \label{eq:chamindetector2}
   && S_{\phi i} \!=\! \int \! \dd^4x \sqrt{-g} \Big[ \beta_i\frac{\phi}{\mpl}T_i \!+\!\nonumber \\ &&c_i^{\mu\nu}(\varphi_0)\frac{\partial_\mu\varphi\partial_\nu\varphi}{M^4}T_i \!+\! \frac{1}{M^4}T_i^{\mu\nu}{{d_i}^{\rho\sigma}}_{\mu\nu}(\varphi_0)\partial_\rho\phi\partial_\sigma\phi \Big] \,,
\end{eqnarray}
where $\mpl$ is the reduced Planck mass, $M$ is the UV-cutoff of the theory, $T_i^{\mu\nu}$ is the energy-momentum tensor for SM particle species $i$, with $T_i=g_{\mu\nu}T^{\mu\nu}_i$ its trace, and with background-dependent tensors $c_i^{\mu\nu}(\varphi_0)$ and $d_{\mu\nu\rho\sigma}(\varphi_0)$. Here we assume for simplicity's sake that $c_i^{\mu\nu}(\varphi_0)\propto \eta^{\mu\nu}$ and ${d_i}^{\mu\nu\rho\sigma}(\varphi_0)\propto \eta^{\mu\rho}\eta^{\nu\sigma}$, where the proportionality coefficients are assumed to be density-dependent and species-dependent constants. As a result we shall concentrate on the following action
\begin{align}
    \label{eq:DEEFT1}
    S&=\int\dd^4 x\sqrt{-g}\left[-\frac12Z^{\mu\nu}(\varphi_0)\partial_\mu\phi\partial_\nu\phi\nonumber\right.\\&\left.-\frac12m^2(\varphi_0)\phi^2-\beta_\gamma\frac{\phi}{\mpl}F_{\mu\nu}F^{\mu\nu}\right.\nonumber\\&\left.+\!\sum_i\!\!\left(\!\beta_i(\varphi_0)\frac{\phi}{\mpl}T_i\!+\!c_i(\varphi_0)\frac{X}{M^4}T_i\!+\!\frac{d_i(\varphi_0)}{M^4}T_i^{\mu\nu}\partial_\mu\phi\partial_\nu\phi\right)\!\!\right],
\end{align}
where $X=-g^{\mu\nu}\partial_\mu\phi\partial_\nu\phi$ and $F^{\mu\nu}$ is the photon field-strength tensor. The sum in Eq.~(\ref{eq:DEEFT1}) includes photons, but since their energy-momentum tensor is traceless they do not contribute to the first two terms. A direct coupling to photons through their field-strength tensor can arise through quantum anomalies~\cite{Brax:2010uq}, and we have therefore included this coupling with strength $\beta_\gamma$. This term is of critical importance since it allows for the production of scalars in the solar tachocline. The couplings to matter arise from the Jordan frame metric, i.e.\ the metric coupling SM particles to the scalar when the graviton is canonically normalised in the underlying covariant theory:
\begin{eqnarray}
    g_{\mu\nu}^J&=&\left ( 1+ 2\beta_i(\varphi_0)\frac{\phi}{\mpl}+2c_i(\varphi_0)\frac{X}{M^4} \right )g_{\mu\nu}\nonumber \\
    &&+ 2\frac{d_i(\varphi_0)}{M^4}\partial_\mu\phi\partial_\nu\phi\,.
    \label{eq:jordaneinstein}
\end{eqnarray}
From a quantum field theory perspective, it is unlikely that the couplings to each matter species are universal, hence our choice to treat the couplings $\beta_i$, $c_i$, and $d_i$ as being species-specific. The term multiplying $g_{\mu\nu}$ is referred to as the \textit{conformal factor} and the term multiplying $\partial_\mu\phi\partial_\nu\phi$ is referred to as the \textit{disformal factor}. We therefore refer to the $\beta_i$s and $c_i$s as conformal couplings, and $d_i$s as disformal couplings ($c_i$ can also be referred to as kinetic-conformal coupling).

The effective theory in Eq.~(\ref{eq:DEEFT1}) includes environmental variations of the various coupling constants via their dependence on the background field $\varphi_0$, and therefore on the local matter density when the energy-momentum tensor of matter is dominated by non-relativistic species, such as in the late-time Universe after matter-radiation equality, or in astrophysical situations through the virialised matter density. In particular, we have allowed the kinetic matrix $Z^{\mu\nu}(\varphi_0)$ to be non-diagonal and to depend on the background field. This structure typically emerges from ghost-free higher-derivative couplings of the field to curvature tensors, and gives rise to the Vainshtein~\cite{Vainshtein:1972sx} and K-mouflage~\cite{Babichev:2009ee} screening mechanisms. We have also allowed the mass to be background field-dependent. This gives rise to the chameleon mechanism~\cite{Khoury:2003aq, Khoury:2003rn}. Note that $m(\varphi_0)$ is the mass of fluctuations about $\varphi_0$. The mass of $\varphi$ in the cosmological background is instead $\mathcal{O}(H_0)$, so that $\varphi$ can act as a DE scalar. Finally, we have allowed the coupling constants $\beta_i$, $c_i$, and $d_i$ to be field-dependent too. This field-dependence is utilised by the symmetron~\cite{Hinterbichler:2010es} and dilaton~\cite{Brax:2010gi} mechanisms.

In this work, we shall focus on theories that utilise the chameleon mechanism, and therefore set $Z^{\mu\nu}(\varphi_0)=\eta^{\mu\nu}$, while taking $\beta_i$, $c_i$, and $d_i$ to be background field-independent. We also introduce the energy scales $M_i \equiv M/d_i^{1/4}$, which depend on the species being considered ($i$). Applying these simplifications to the action in Eq.~(\ref{eq:DEEFT1}), the effective theory considered is:
\begin{align}
    \label{eq:DEEFT2}
    S&=\int\dd^4 x\sqrt{-g}\left[-\frac12\partial_\mu\phi\partial^\mu\phi\nonumber\right.\\&\left.-\frac12m^2(\varphi_0)\phi^2-\beta_\gamma\frac{\phi}{\mpl}F_{\mu\nu}F^{\mu\nu}\right.\nonumber\\&\left.+\sum_i\left(\beta_i\frac{\phi}{\mpl}T_i+c_i\frac{X}{M^4}T_i+\frac{1}{M_i^4}T_i^{\mu\nu}\partial_\mu\phi\partial_\nu\phi\right)\right].
\end{align}

The mechanism by which DE is produced in the Sun and either scatters or is absorbed in DM direct detection experiments is the following. The Sun is necessarily screened, implying that the mass of the scalar in the Sun is $m_\odot=m(\varphi_\odot)>10^3H_0$~\cite{Brax:2011aw,Wang:2012kj}. In practice, the high density of the Sun with respect to the cosmological background ($\rho_\odot/\rho_c \sim 10^{29}$) implies that $m_\odot$ is in fact much heavier than this deep inside the Sun. The exact value is model-dependent, as it depends on the exact density dependence of the scalar's mass. The high mass prevents the efficient production of chameleons via Compton or bremsstrahlung processes in the Sun's core via a Boltzmann suppression. However, the direct coupling to photons allows for production in the magnetic field of the tachocline via a mechanism analogous to the Primakoff process for axions~\cite{Primakoff:1951pj}: we review the chameleon production process in Appendix~\ref{sec:production}. The relevant operators for this production process are:
\begin{align}
    {S}_{\phi\gamma} =&\int \dd^4x \sqrt{-g}\bigg[-\frac{1}{4}F_{\mu\nu}F^{\mu\nu} - \beta_\gamma\frac{\phi}{\mpl}F_{\mu\nu}F^{\mu\nu}\nonumber\\ &+\frac{1}{M_\gamma^4}T^{\mu\nu}_{\gamma}\partial_\mu \phi \partial_\nu \phi
     \bigg]\,,\label{eq:LagrangianEM}
\end{align}
where $M_\gamma={M}/{d_\gamma^{1/4}}$ is the energy scale related to the disformal coupling to photons. As chameleons are produced in the tachocline and not in the core, we need to impose that the mass of the chameleon in the core be larger than the local temperature. This can be achieved using the density-dependence of the mass, as the ratio between the densities in the core and in the tachocline is around two orders of magnitude.

Once produced, solar chameleons free-stream out of the Sun, with a fraction passing through the Earth, and an even smaller fraction through the chambers of DM direct detection experiments. In these chambers the mass is mostly determined by the detector's geometry~\cite{Khoury:2003aq,Khoury:2003rn} and is typically small ($m_{\rm DC} \ll m_e$, where $m_e$ is the mass of the electron). Chameleons can therefore be treated as massless particles in the chambers of DM direct detection experiments for all intents and purposes. Chameleons passing through the detector chamber can scatter off or be absorbed by the particles utilised for the detection, via couplings of the form
\begin{equation}
    \label{eq:chamindetector}
    S_{\phi i} \!=\! \int \! \dd^4x \sqrt{-g}\left[ \beta_i\frac{\phi}{\mpl}T_i \!+\! c_i\frac{X}{M^4}T_i \!+\! \frac{1}{M_i^4}T_i^{\mu\nu}\partial_\mu\phi\partial_\nu\phi\right ] \,.
\end{equation}
These couplings give rise to what we refer to as the ``chameleo-electric effect'', and correspondingly to electron recoils in the ${\cal O}({\rm keV})$ range. Similar interactions with neutrons and protons will instead give rise to atomic recoils, which we will not explicitly consider in this paper.

In the subsequent analysis, we will neglect the coupling $c_i$, which controls the strength of the kinetic-conformal coupling $XT_i$. We find that this coupling has an effect similar to that of the conformal coupling $b_i$ at the level of detection. Moreover, from a statistical perspective, it does not lead to a substantial improvement in the fit to the XENON1T signal (which would otherwise warrant its inclusion as a free parameter), as the disformal detection channel dominates over the conformal one(s), for reasons we will  discuss in Sec.~\ref{sec:analysis} below. The choice of setting $c_i=0$ for the purposes of this work should be viewed as a simplifying working assumption, which sets a minimal phenomenologically working model: (re)-including $c_i$ would not change our results, nor the goodness of the fit.

\section{Model}
\label{sec:model}

\subsection{Theoretical considerations}

As it stands, our effective theory is still too general to calculate the production and detection processes because we need to determine the free functions $m^2(\phi_0)=m^2(\vec{x})=m^2(\rho)$, $\beta_i(\rho)$, and $d_i(\rho)$. There are two possibilities for fixing the spatial-dependence of the scalar's mass. The first is to parameterise our ignorance by assuming a functional form for its density and constraining the parameters associated with this choice. The second is to provide an {explicit}  model and thereby to calculate the spatial-dependence explicitly. We opt for the second choice for three reasons. First, it is not guaranteed that an arbitrary fitting function will reproduce the dynamics of any fundamental theory, so it is not clear what information is lost by making such a choice. Second, chameleon models are well-constrained by laboratory and astrophysical probes, so choosing a well-studied model allows us to explore complementarity with these bounds, and to determine the feasibility of our scenario by identifying the existence of regions of parameter space where our model can simultaneously satisfy these bounds and successfully explain the XENON1T signal. Finally, chameleons have $\beta_i$ and $d_i$ constant, so we can exemplify our scenario using a simple minimal model.

Chameleon theories are subject to a  no-go theorem~\cite{Wang:2012kj} that excludes the possibility of \emph{self-acceleration} defined strictly as acceleration in the Jordan frame but not the Einstein frame in the complete absence of any cosmological constant. Dark energy scenarios where the acceleration is driven by the scalar potential, i.e.\ a quintessence-like explanation, are not excluded by this theorem, but require a tuning of an overall additive constant. Of course, such a tuning is also present in self-acceleration scenarios as this additive constant is arbitrarily set to zero. The generalized couplings (disformal and kinetic-conformal, $XT$) that we consider here are additional potential caveats to the no-go theorem since they were not considered when deriving it~\cite{Noller:2012sv}. Other caveats are discussed in  Ref.~\cite{Wang:2012kj}.

As discussed in more detail in Appendix~\ref{sec:AppendixA}, the density-dependence of the chameleon's mass arises because its dynamics are governed by an effective potential
\begin{equation}
    \label{eq:effectivepotential_main}
    V_{\rm eff} (\phi)= V(\phi) + \rho \exp\left(\frac{\beta_m\phi}{M_{\rm Pl}}\right)\,,
\end{equation}
where $\rho$ is the density of the matter species coupled to the chameleon, $\beta_m$ is the strength of such coupling, and $V(\phi)$ is the bare potential, which would govern the dynamics if the field were not coupled to matter. A common model for the bare potential is the  power-law {chameleon~\cite{Khoury:2003aq,Khoury:2003rn,Brax:2004qh} leading to a density-dependent mass at the minimum $\phi_{\rm min}$ of the effective potential
\begin{eqnarray}
\label{eq:mphi_main}    
m_\phi^2 &=& \frac{{\mathrm d}^2V_{\rm eff}(\phi)}{{\mathrm d}\phi^2}\bigg|_{\phi = \phi_{\rm min}} \!\! \nonumber\\
&=& n( 1+n)\Lambda^{4+n}\left(\frac{\beta_m\,\rho}{n\mpl\Lambda^{4+n}}\right)^{\frac{2+n}{1+n}}\,,
\end{eqnarray}
where $\Lambda$ is an energy scale and $n$ is a power-law index {($V(\phi) \propto \phi^{-n}$)}. Note that both $n>0$ (inverse power-law) and $n<0$ (power-law) can lead to the chameleon behaviour provided $n\ne -1,\,-2$, or an odd negative integer.  
We also assume that $\beta_m\phi/M_{\rm Pl} \ll 1$, in order for the excursion of the chameleon field not to exceed $M_{\rm Pl}/\beta_m$. Note that the swampland conjectures (see Ref.~\cite{Obied:2018sgi}) lead to a {lower} bound on $\beta_m \gtrsim c V/\rho$, where $c$ is a constant of order unity~\cite{Brax:2019rwf}.}

\subsection{Production in the Sun}

Chameleons can be resonantly produced in a dense magnetised plasma when the chameleon mass matches the plasma frequency of the environment. This process, governed by the chameleon-photon coupling in Eq.~\eqref{eq:LagrangianEM}, occurs in the Sun within a narrow shell whose location depends on the chameleon rest mass~\cite{Brax:2011wp, Brax:2015fya}. Chameleon production can also occur through non-resonant processes, which occur in all magnetised regions inside the Sun. Here we adopt the non-resonant production mechanism and consider a magnetic field profile $B = B(r)$, where $r$ is the radial coordinate.~\footnote{Note that we do not consider couplings of chameleons to longitudinal plasmons, as recently considered in Refs.~\cite{Caputo:2020quz,OHare:2020wum}.} For the solar model, we adopt the profiles described in Ref.~\cite{Bahcall:2004fg}, which has also been used to derive the formula for the Primakoff flux used by the XENON1T collaboration, see Ref.~\cite{Kuster:2008zz}. We also note that there is some disagreement in the field between different solar models, see e.g.\ Refs.~\cite{Asplund:2009fu, Serenelli:2009yc, Villante:2013mba, Vagnozzi:2016cmr, Serenelli:2016nms, Vagnozzi:2017wge}.

The resulting differential flux per unit energy of solar chameleons on Earth, resulting from isotropic production in the Sun, is given by
\begin{equation}
    \label{eq:Earthflux}
    \frac{\mathrm{d}\Phi_{\rm Earth}}{\mathrm{d}\omega} = \frac{\pi R_t^2}{4\pi d_\odot^2}\frac{\mathrm{d}\Phi}{\mathrm{d}\omega}\,,
\end{equation}
where $d_\odot = 1\,$A.U.\ is the Earth-Sun distance, and $R_t \sim 0.7R_{\odot}$ is the tachocline radial coordinate. The flux of chameleons produced in the Sun, $\dd\Phi/\dd\omega$, is calculated in Appendix~\ref{sec:production}, see Eq.~\eqref{eq:production1}.

In principle, one could also consider production from toroidal magnetic modes deeper within the Sun. However, we note that there are significant uncertainties associated to the strength and profile of these modes~\cite{2009LRSP....6....4F,2010LRSP....7....3C, 2016Natur.535..526W}. The equipartition value for the large-scale toroidal magnetic field due to shearing is $\sim 1\,{\rm T}$~\cite{2009LRSP....6....4F,2010LRSP....7....3C,2016Natur.535..526W}. As we later assume a strength $B_t=30\,{\rm T}$ for the tachocline magnetic field, we expect the associated contribution to the chameleon flux relative to the contribution we considered to be suppressed by a factor $(1/30)^2 \approx {\cal O}(10^{-3})$, and to appear at higher energies than those of interest for the XENON1T excess. While the strength of these modes may be amplified locally by up to ${\cal O}(10^2)$ their equipartition values in the convection zone to form sunspots~\cite{2009LRSP....6....4F}, the significant uncertainties at play prevent us from fully quantifying the impact of these processes on our results. We thus conservatively choose to neglect the effect of toroidal magnetic modes on chameleon production deeper within the Sun, noting that these could lead to subdominant features in the event rate at higher energies, but deferring a full study to follow-up work.

We have not considered other more ``standard'' production mechanisms, which in the parameter space spanned are subdominant. For example, chameleons could be produced through so-called ABC reactions (atomic recombination and deexcitations, bremsstrahlung, and Compton), in a similar fashion to axions~\cite{Raffelt:1985nk, Dimopoulos:1986kc, Redondo:2013wwa}. However, by virtue of the chameleon mass being density-dependent, we can always find large regions of parameter space where these processes are kinematically disfavoured (in particular by adjusting the energy scale $\Lambda$). As discussed at the end of Appendix~\ref{sec:AppendixA}, this can be achieved by requiring that $m_{\rm eff}^2(\rho_{\rm core}) \gtrsim T_{\rm core}^2$, where $m_{\rm eff}^2 \equiv m_\phi^2- \omega^2_{\rm Pl}$, with $\omega^2_{\rm Pl}$ the plasma frequency squared given by Eq.~\eqref{plas}, $\rho_{\rm core} \simeq 150\,{\rm g}/{\rm cm}^3$ the Sun's core density, and $T_{\rm core} \simeq 1.5 \times 10^7\,{\rm K}$ the Sun's core temperature. Typically we expect that the mass of the chameleons scale like $m(\rho)\simeq \rho^{\alpha}$ where $\alpha=3/4$ for chameleons with $n=1$. This implies that the mass  deep inside the core increases typically by two orders of magnitude compared to the mass in the tachocline. If this condition is satisfied, production of solar chameleons in the deeper regions of the Sun is kinematically forbidden, with the overall flux being dominated by Primakoff-like production in the tachocline.

From Eq.~(\ref{eq:mphi_main}), we see that the previous condition translates into constraints on $\beta_m$, $\Lambda$, and $n$. Fixing $\beta_m$ and $n>0$, we find that this condition sets an upper limit on the allowed value of $\Lambda$. Later in our analysis we will consider $\beta_m \simeq 10^2$ and $n=1$, since we find that the event rate in XENON1T is mostly sensitive to $\beta_{\gamma}$ and $M_e$, and only weakly sensitive to $\beta_m$, $n$, $\Lambda$, and $M_{\gamma}$. For this choice of $\beta_m$ and $n$, we find that production of solar chameleons in the Sun's core is kinematically forbidden as long as $\Lambda \lesssim 1\,\mu{\rm eV}$. We remark again that fixing $\Lambda$ to other values has no appreciable effect on the XENON1T event rate, and hence on our results.

\subsection{Chameleon detection}

To reach the XENON1T detector and leave detectable imprints, solar chameleons with energies $\omega \gtrsim {\cal O}({\rm keV})$ need to propagate through various dense media unscathed. In general, chameleons with incoming energy $\omega$ will be able to traverse a dense barrier of energy $\rho$ provided $\omega \gtrsim m(\rho)$. Let us focus on the benchmark point we discussed above and which we consider throughout the paper, where $\beta_m \simeq 10^2$, $\Lambda \lesssim 1\,\mu{\rm eV}$, and $n=1$, so that $m \propto \rho^{3/4}$. We thus need to ensure that chameleons make it through the densest material along their path. The highest density involved in the problem is that of lead, which the XENON1T detector is partially made of, and whose density is $\rho_{\rm Pb} \sim 10\,{\rm g}/{\rm cm}^3$. For the above parameters, we find $m(\rho_{\rm Pb}) \sim 0.6\,{\rm keV}$, meaning that chameleons with energies $\omega \gtrsim 0.6\,{\rm keV}$ are able to reach the XENON1T detector. This is sufficient to ensure that chameleons are able to fit the XENON1T excess, which occurs at energies higher than this cut-off. Moreover, as rocks and the tachocline have densities respectively one and two orders of magnitude lower than that of lead, this also ensures that chameleons are able to escape the tachocline and propagate through the rock which makes up Gran Sasso (being mostly made of limestone, with density $\rho \sim 3\,{\rm g}/{\rm cm}^3$).

In the XENON1T detector, solar chameleons can be absorbed by electrons via the chameleo-electric effect. This is the chameleon analogue of the photoelectric and axio-electric effects for photons and axions respectively. The cross-section for the chameleo-electric effect in DM direct detection experiments, $\sigma_{\phi e}$, is computed in Appendix~\ref{sec:detectioncrosssection}, see in particular Eq.~(\ref{eq:chameleoelectric}). The resulting differential event rate per unit production energy $\omega$ for chameleon absorption by electrons in the XENON1T detector is given by:
\begin{equation}
    \label{eq:drdomegath}
    \left ( \frac{\mathrm{d}R}{\mathrm{d}\omega} \right )_{\rm th} = \epsilon(\omega)\,\int \frac{\mathrm{d}R_0(\omega_R)}{\mathrm{d}\omega_R}\,\Theta(\omega-\omega_R)\,\mathrm{d}\omega_R\,,
\end{equation}
where $\Theta(\omega)$ is the energy resolution of the detector and $\epsilon(\omega)$ is the XENON1T detection efficiency, given in Fig.~2 of Ref.~\cite{Aprile:2020tmw}.\footnote{As described below Eq.~(1) in Ref.~\cite{Aprile:2020tmw}, the efficiency does not enter the integral in the convolution function.} The ``raw'' differential event rate per unit production energy of chameleons in the XENON1T detector, i.e.\ not taking into account energy resolution and detection efficiency effects, is given by:
\begin{equation}
    \label{eq:diffrate_xenon}
    \frac{\mathrm{d}R_0(\omega)}{\mathrm{d}\omega} = N_{\rm Xe}\,\frac{\mathrm{d}\Phi_{\rm Earth}}{\mathrm{d}\omega}\,\sigma_{\phi e}\,,
\end{equation}
where the expression for the flux at Earth is given in Eq.~\eqref{eq:Earthflux}, and where the number of atoms per ton of xenon is $N_{\rm Xe} = 4.6\times 10^{27}{\rm \,ton}^{-1}$, so that the differential event rate is expressed in units of ${\rm ton}^{-1}\,{\rm yr}^{-1}\,{\rm keV}^{-1}$. We have appended the subscript $_{\rm th}$ to the differential event rate per production energy in Eq.~(\ref{eq:drdomegath}) to stress that this is a \textit{theoretical} event rate, which depends on the underlying chameleon parameters through the dependence of $\mathrm{d}\Phi_{\rm Earth}/\mathrm{d}\omega$ and $\sigma_{\phi e}$ in Eq.~(\ref{eq:diffrate_xenon}) on these parameters. Comparing the theoretical event rate against the event rate \textit{measured} by XENON1T will allow us to set constraints on the underlying chameleon parameters.

While we focus on the XENON1T apparatus, we stress that the results in Appendix~\ref{sec:production} and Appendix~\ref{sec:detectioncrosssection}, from which we derive Eqs.~\eqref{eq:drdomegath}-\eqref{eq:diffrate_xenon}, are more broadly applicable. In particular, they can be applied to future DM direct detection experiments, which we discuss in more detail in Sec.~\ref{subsec:other}. Note, moreover, that because we used an effective action, the expression for the cross-section we compute, as well as its derivation, are general results that can be applied to any mass and coupling and for different detector setups. Our goals are now to explore whether solar chameleons are able to account for at least part of the observed low-recoil excess observed in XENON1T and, if so, to determine benchmark examples of solar chameleon parameters which provide an adequate fit to the XENON1T signal.

As discussed in Sec.~\ref{sec:chameleons}, we do not include the kinetic-conformal $c_i$ coupling (i.e.\ the term proportional to $XT_i$), since we find \emph{a posteriori} that including this operator does not lead to a substantial improvement in fit to the XENON1T signal, which would instead be required to warrant its inclusion as a free parameter. If we were to include this term, a fit to XENON1T data would strongly prefer setting the associated coupling to zero, leading to the one-parameter extension not being preferred from a statistical point of view. This may not be the case for other DE models but it is for the specific case of chameleon DE.

The physical reason why this coupling worsens the fit to the XENON1T signal is that the associated cross-section does not scale fast enough with energy $\omega$. This leads to the corresponding peak in the resulting event rate being below the $\approx 2\,{\rm keV}$ required to explain the XENON1T excess. On the other hand, the disformal coupling leads to additional powers of $\omega$ in the associated cross-section, thereby moving the peak to the correct position  to explain the XENON1T excess.

\section{Analysis}
\label{sec:analysis}

In principle, the parameter space describing production in the Sun and subsequent detection in the XENON1T detector of solar chameleons is six-dimensional, and spanned by the following parameters: the coupling to matter (in this case electrons) $\beta_e \equiv \beta_m$, the scale governing the disformal coupling to electrons (hereafter ``electron disformal scale'') $M_e=M/d_e^{1/4}$, the coupling to photons $\beta_{\gamma}$, the scale governing the disformal coupling to photons (hereafter ``photon disformal scale'') $M_{\gamma}=M/d_\gamma^{1/4}$, and finally the energy scale $\Lambda$ and power-law index $n$ describing the chameleon self-interaction potential as given in Eq.~(\ref{eq:potential}).

To simplify our analysis, we set $\Lambda=1\,\mu{\rm eV}$ and $n=1$. As explained earlier, requiring $\Lambda \lesssim {\cal O}(\mu{\rm eV})$ ensures that the chameleon's effective mass is sufficiently high in the core of the Sun so that production of chameleons through the usual Compton and bremsstrahlung mechanisms is kinematically suppressed. On the other hand, $n=1$ corresponds to the best-studied chameleon model, with bounds typically only being reported for this specific choice~\cite{Burrage:2016bwy}. These extensive studies have excluded a large region of parameter space~\cite{Burrage:2017qrf}, and it is thus of interest to explore whether XENON1T is able to probe part of the remaining parameter space of this model. In any case, we have explicitly verified that fixing $\Lambda$ and $n$ to other values affects chameleon production and the resulting event rate well below the $\%$-level, and thus has no appreciable effects on our results.

These choices leave us with 4 parameters: $\beta_e$, $M_e$, $\beta_{\gamma}$, and $M_{\gamma}$. However, we anticipate that XENON1T will be mostly sensitive to $\beta_{\gamma}$ and $M_e$, and very weakly sensitive to $\beta_e$ and $M_{\gamma}$, for the following reasons. Firstly, we expect the disformal detection channel to dominate over the conformal one, as the former scales with a higher power of energy than the latter, see Eq.~(\ref{eq:chameleoelectric}). This feature moves the peak in the event rate towards higher energies, better fitting the XENON1T excess. Therefore, detection in XENON1T is mostly controlled by the electron disformal scale $M_e$ rather than the coupling $\beta_e$, with the associated cross-section scaling as $1/M_e^8$, see Eq.~(\ref{eq:chameleoelectric}). Second, if we require that $M_{\gamma} \gg {\cal O}({\rm keV})$ so that Primakoff production in horizontal branch stars does not dominate over neutrino losses~\cite{Brax:2014vva}, production in the Sun will predominantly proceed through the conformal channel.\footnote{Note that the bounds on $M_{\gamma}$ derived in Ref.~\cite{Brax:2014vva} do not directly apply to our scenario, as they were derived assuming that the scalar is massless.} As a result, production will mostly be controlled by the photon coupling $\beta_{\gamma}$ rather than the photon disformal scale $M_{\gamma}$. In particular, the associated production flux scales as $\beta_{\gamma}^2$, see Eq.~(\ref{eq:production1}). Finally, the expected event rate in XENON1T, which is the only quantity we can directly compare to observations, depends only on the product of the production flux and the detection cross-section, as can be clearly seen in Eq.~(\ref{eq:diffrate_xenon}). This product scales as $\beta_{\gamma}^2/M_e^8$. We therefore expect that the XENON1T measurements will predominantly constrain the following parameter combination, which we denote by $\beta_{\rm eff}$, and refer to as the effective coupling:
\begin{eqnarray}
\label{eq:betaeff}
\beta_{\rm eff} \equiv \beta_{\gamma} \left ( \frac{\rm keV}{M_e} \right )^4 \,.
\end{eqnarray}
We can view $\beta_{\rm eff}$ as being the chameleon equivalent of the product $g_{a\gamma}g_{ae}$ for the case of solar axions produced via the Primakoff effect in the Sun and detected via the axio-electric effect in the XENON1T detector. In particular, the expected event rate in the XENON1T detector is proportional to $\beta_{\rm eff}^2$, as we derive in Eq.~\eqref{eq:diffrate_xenon0} in the appropriate limit discussed in the Appendix.

We now proceed to analyze the XENON1T measurements to verify whether solar chameleons can fit these measurements, and whether our previous expectations are met. We perform a Bayesian statistical analysis to constrain the four chameleon parameters, which we collectively denote by $\boldsymbol{\theta} \equiv \{ \beta_{\gamma}\,,M_{\gamma}\,,\beta_e\,,M_e \}$. The likelihood ${\cal L}(\boldsymbol{\theta} \vert \boldsymbol{d})$ to observe the data $\boldsymbol{d}$ given a certain set of model parameters $\boldsymbol{\theta}$ is
\begin{eqnarray}
	\label{eq:likelihood}
	{\cal L}(\boldsymbol{\theta} \vert \boldsymbol{d}) = \exp \left [ -\frac{\chi^2(\boldsymbol{\theta},\boldsymbol{d})}{2} \right ]\,,
\end{eqnarray}
where the $\chi^2$ function entering the likelihood is given by 
\begin{eqnarray}
	\label{eq:chi2}
	\chi^2(\boldsymbol{\theta},\boldsymbol{d}) \!=\! \sum _i\! \left [ \frac{(\frac{\mathrm{d}R}{\mathrm{d}\omega})_{\rm th}(\boldsymbol{\theta}) \!+\! B_0 \!-\! (\frac{\mathrm{d}R}{\mathrm{d}\omega})_{\rm meas}(\boldsymbol{d})}{\sigma_i^2} \right ]^2\,,
\end{eqnarray}
with the sum being performed over the energy bins $\omega_i$ at which XENON1T measure their event rates. In Eq.~(\ref{eq:chi2}), $(\mathrm{d}R/\mathrm{d}\omega)_{\rm th}$ denotes the theoretical event rate given by Eq.~(\ref{eq:drdomegath}), $(\mathrm{d}R/\mathrm{d}\omega)_{\rm meas}$ denotes the rate as measured by XENON1T (black re-binned datapoints of Fig.~4 in Ref.~\cite{Aprile:2020tmw}), and $B_0$ denotes the background model (red curve of Fig.~4 in Ref.~\cite{Aprile:2020tmw}), which is itself a function of energy. The XENON1T background model is described in more detail in Sec.~IIIB of Ref.~\cite{Aprile:2020tmw} (see in particular their Table~1 and Fig.~3), and includes contributions from ten different components, ranging from $^{214}$Pb to solar neutrinos. We refer the reader to Ref.~\cite{Aprile:2020tmw} for more a detailed discussion of $B_0$.

For simplicity, we only consider the first 16 bins, in the recoil energy range $1.5\,{\rm keV} \lesssim \omega_R \lesssim 16.5\,{\rm keV}$. We do not consider bins beyond the 16$^{\rm th}$ for two reasons:
\begin{enumerate}
\item the theoretical solar chameleon event rate drops quickly beyond the third bin, partly due to the limited width of the differential chameleon flux (see Fig.~\ref{fig:fig2} below), and to the effects of energy resolution and detector efficiency, entering in Eq.~(\ref{eq:drdomegath}) through $\Theta(\omega)$ and $\epsilon(\omega)$;
\item the measured rate in the bins beyond the third is highly consistent with the background model $B_0$, and therefore does not call for new physics explanations: the only exceptions are a few anomalous bins (the 17$^{\rm th}$, the 20$^{\rm th}$, the 24$^{\rm th}$, the 26$^{\rm th}$, and the 29$^{\rm th}$ bins respectively), which none of the proposed theoretical models (including the solar axions, neutrino magnetic moment, and tritium explanations invoked by the XENON1T collaboration) have been able to explain.\footnote{Note that also the 11$^{\rm th}$ and 14$^{\rm th}$ bins lie $\approx 1\sigma$ above the background model $B_0$. The solar axion explanation of the low-energy excess improves the fit to the 14$^{\rm th}$ bin through the contribution of $^{57}$Fe axions, see Fig.~7b in Ref.~\cite{Aprile:2020tmw}.}
\end{enumerate}
Of these 16 bins, the second and third deviate the most from the XENON1T background model $B_0$, and therefore contribute the most to the excess.

We impose flat priors on $\log_{10}\beta_{\gamma} \in [0; 11]$, $\log_{10}(M_{\gamma}/{\rm keV}) \in [0; 25]$, $\log_{10}\beta_e \in [1; 2]$, and $\log_{10}(M_e/{\rm keV}) \in [0; 25]$. For $\beta_{\gamma}$, the upper prior edge is motivated by the latest results from the Kinetic Weakly Interacting Slim Particles (KWISP) detector on the CAST axion search experiment at CERN, which finds $\beta_{\gamma}<10^{11}$~\cite{Cuendis:2019mgz}, whereas the lower prior edge is arbitrary (we have checked that extending it to lower values does not qualitatively affect our final results). Due to the weak sensitivity of the XENON1T measurements to $\beta_e$, we have chosen a narrow prior for this parameter. We have explicitly verified that extending the prior further has no effects on our results. Similarly, fixing $\beta_e$ (for instance to $\beta_e=10^2$) would also have no effect on our results, see later discussion below Eq.~(\ref{eq:betaeffbestfit}).

We allow the photon and electron disformal scales to span the range between the ${\rm keV}$ scale and the Planck scale. It is worth noting that limits on the disformal scale $M$ obtained from collider searches, indicating $M \gtrsim {\cal O}(100\,{\rm GeV})$ as for instance in Refs.~\cite{Brax:2015hma, Brax:2016did, Aaboud:2019yqu,Trojanowski:2020xza} (including works from one of us), only apply to the chameleon-quark disformal coupling scale, and not to the scale governing the coupling to photons and electrons. The strongest bound on the scalar-photon disformal coupling comes from demanding that Primakoff production of scalars in horizontal branch stars does not dominate over neutrino losses, and the strongest bound on the scalar-electron disformal coupling similarly derives from demanding that losses from Compton and bremsstrahlung production do not significantly alter the properties of these objects. In both cases, the bounds impose $M_e,\, M_{\gamma}\gtrsim \mathcal{O}(0.1\,{\rm GeV})$ as derived by one of us in Ref.~\cite{Brax:2014vva}. Note, however, that these bounds only apply in the limit where the scalar's mass can be neglected. This is not the case in our scenario since we impose $m_{\rm eff}(\varphi_\odot)>T_\odot$, where $T_\odot$ is the core temperature of the Sun. The bounds derived  in Ref.~\cite{Brax:2014vva} therefore do not directly apply to our case. Moreover, as we will discuss in Sec.~\ref{sec:stellarconstraints}, we expect production within these stellar objects to be kinematically suppressed for the benchmark point in parameter space we considered. We consequently conservatively choose to allow $M_e$ and $M_{\gamma}$ to be as low as ${\cal O}({\rm keV})$, but not any lower. As discussed at the end of Appendix~\ref{sec:production}, for $M_e\,,M_{\gamma} \lesssim {\cal O}({\rm keV})$, the back-reaction effect of the disformal coupling on the scalar field profile in the Sun can become non-negligible, resulting in the break-down of the resulting production flux computation. The ${\cal O}({\rm keV})$ scale which determines whether or not this effect is negligible is set by the maximal temperature reached within the Sun.

To sample the posterior distribution of the chameleon parameters we use Markov Chain Monte Carlo (MCMC) methods. We make use of the cosmological sampler \texttt{Montepython3.3}~\cite{Brinckmann:2018cvx}, configured to act as a generic sampler. The convergence of the generated MCMC chains is monitored through the Gelman-Rubin parameter $R-1$~\cite{Gelman:1992zz}, and we require $R-1<0.01$ for the chains to be considered converged.

Finally, we also quantify the significance of the preference (if any) for the solar chameleon model over the background-only model $B_0$. We do so by adopting the same test statistic $q(s)$ used by the XENON1T collaboration, with $s$ symbolically denoting the signal parameters, see Eq.~(17) in Ref.~\cite{Aprile:2020tmw}. This is essentially a profile log-likelihood test statistic. The statistical significance of the preference (if any) for the solar chameleon model is then determined by $q(0)$, i.e.\ comparing the difference in goodness-of-fit of the best-fit solar chameleon model relative to a $B_0$-only fit.

\section{Results}
\label{sec:results}

We now analyze the XENON1T data using the methodology, priors, and likelihood described in Section~\ref{sec:analysis}. We perform an MCMC run on the four-dimensional parameter space spanned by $\beta_{\gamma}$, $\beta_e$, $M_{\gamma}$, and $M_e$. This MCMC run confirms our earlier expectation that we are only sensitive to the parameter combination of $\beta_{\gamma}$ and $M_e$ given by $\beta_{\rm eff}$ in Eq.~(\ref{eq:betaeff}), while not being sensitive to $\beta_e$ and $M_{\gamma}$. We shall discuss the obtained constraints on $\beta_{\rm eff}$ later on.

\begin{figure}
    \includegraphics[width=1.0\linewidth]{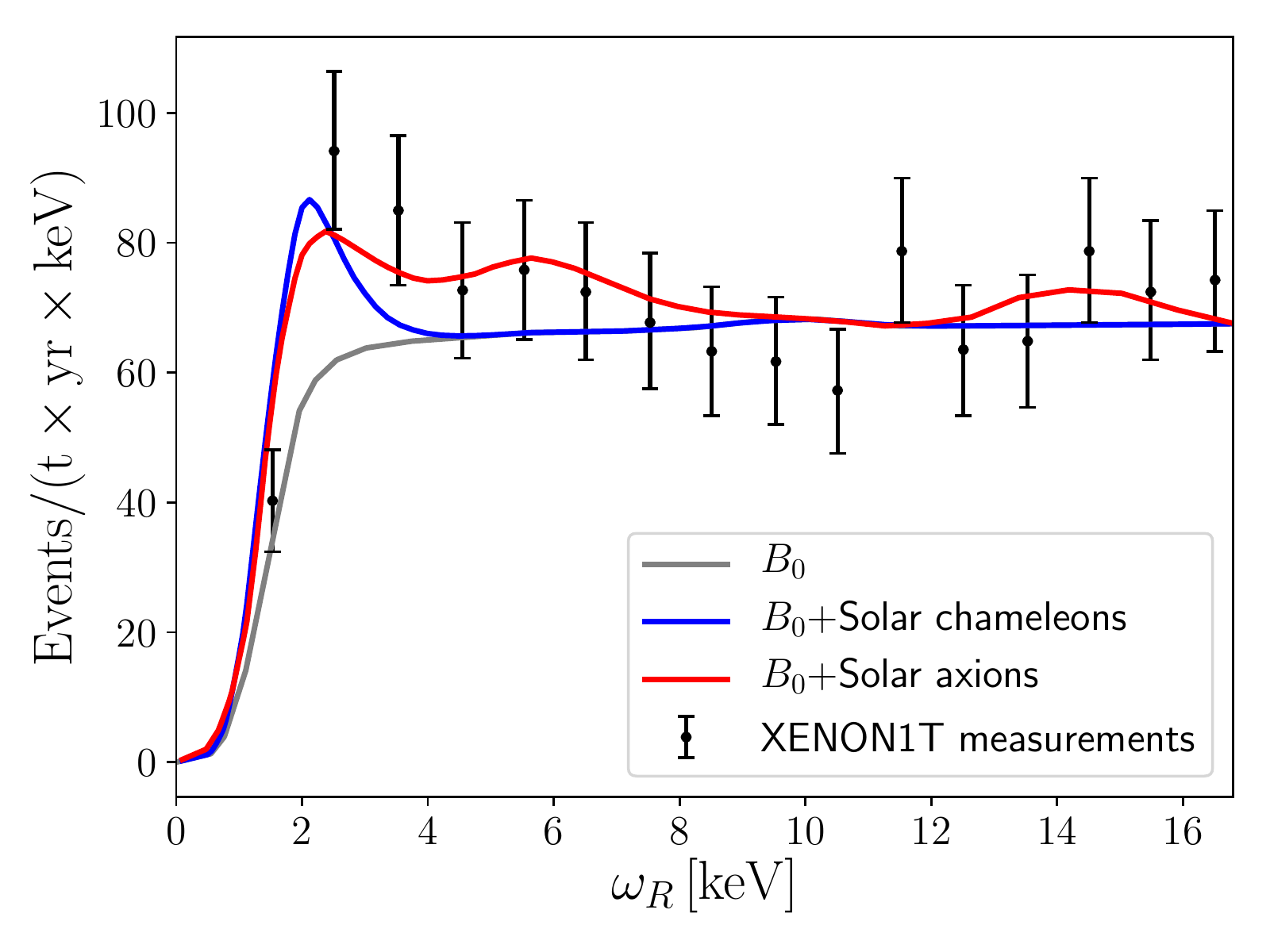}
    \caption{Benchmark example of a solar chameleon fit to the XENON1T signal (event rate in units of ${\rm ton}^{-1}\,{\rm yr}^{-1}\,{\rm keV}^{-1}$ versus recoil energy in units of ${\rm keV}$), representative of the best achievable fit within this scenario. The chameleon parameters are fixed to $\beta_e=10^2$, $M_e=10^{3.6}\,{\rm keV}$, $\beta_{\gamma}=10^{10}$, $M_{\gamma}=1000\,{\rm TeV}$, $\Lambda=1\,\mu{\rm eV}$, and $n=1$. The black data points denote the XENON1T measurements, the grey curve is the XENON1T background model $B_0$, and the blue curve gives the event rate predicted by the solar chameleon model with parameters fixed to the aforementioned values. The fit improves with respect to a background-only fit by $\Delta \chi^2 \simeq -4.0$, with a total $\chi^2=13.2$ for 16 datapoints. Various combinations of the chameleon parameters can lead to a fit of identical quality to the benchmark fit shown here, which is almost entirely governed by the combination of $\beta_{\gamma}$ and $M_e$ given by $\beta_{\rm eff}$ in Eq.~(\ref{eq:betaeff}), see also Eq.~(\ref{eq:betaeffbestfit}). Also shown for comparison (red curve) is the best-fit result for the signal derived from the solar axion model discussed by the XENON1T collaboration in Ref.~\cite{Aprile:2020tmw}, see Fig.~7b therein.}
    \label{fig:fig1}
\end{figure}

A very important result of our analysis is that we are able to identify regions/benchmark points in parameter space which provide a good fit to the XENON1T signal (to be quantified shortly). One such benchmark example is presented in Fig.~\ref{fig:fig1}, where the blue curve is obtained by fixing the chameleon parameters to $\beta_e=10^2$, $M_e=10^{3.6}\,{\rm keV}$, $\beta_{\gamma}=10^{10}$, $M_{\gamma}=1000\,{\rm TeV}$, $\Lambda=1\,\mu{\rm eV}$, and $n=1$. The black data points denote the XENON1T measurements, and the grey curve is the XENON1T background model $B_0$ (the measurements and background are taken from Ref.~\cite{Aprile:2020tmw}). Overall, the resulting fit to the XENON1T signal is good, with a best-fit $\chi^2_{\min} \simeq 13.2$ for 16 datapoints. Moreover, we find an improvement in fit of $\chi^2_{\min}-\chi^2_{B_0} \simeq -4.0$ with respect to a $B_0$-only fit, with the latter delivering $\chi^2_{B_0} \simeq 17.2$. Under this signal model, and using the previously discussed $q(s)$ profile log-likelihood test statistic, a $B_0$-only fit to the signal is rejected at a significance of  $2.0\sigma$.

\begin{figure}[!bt]
    \includegraphics[width=1.0\linewidth]{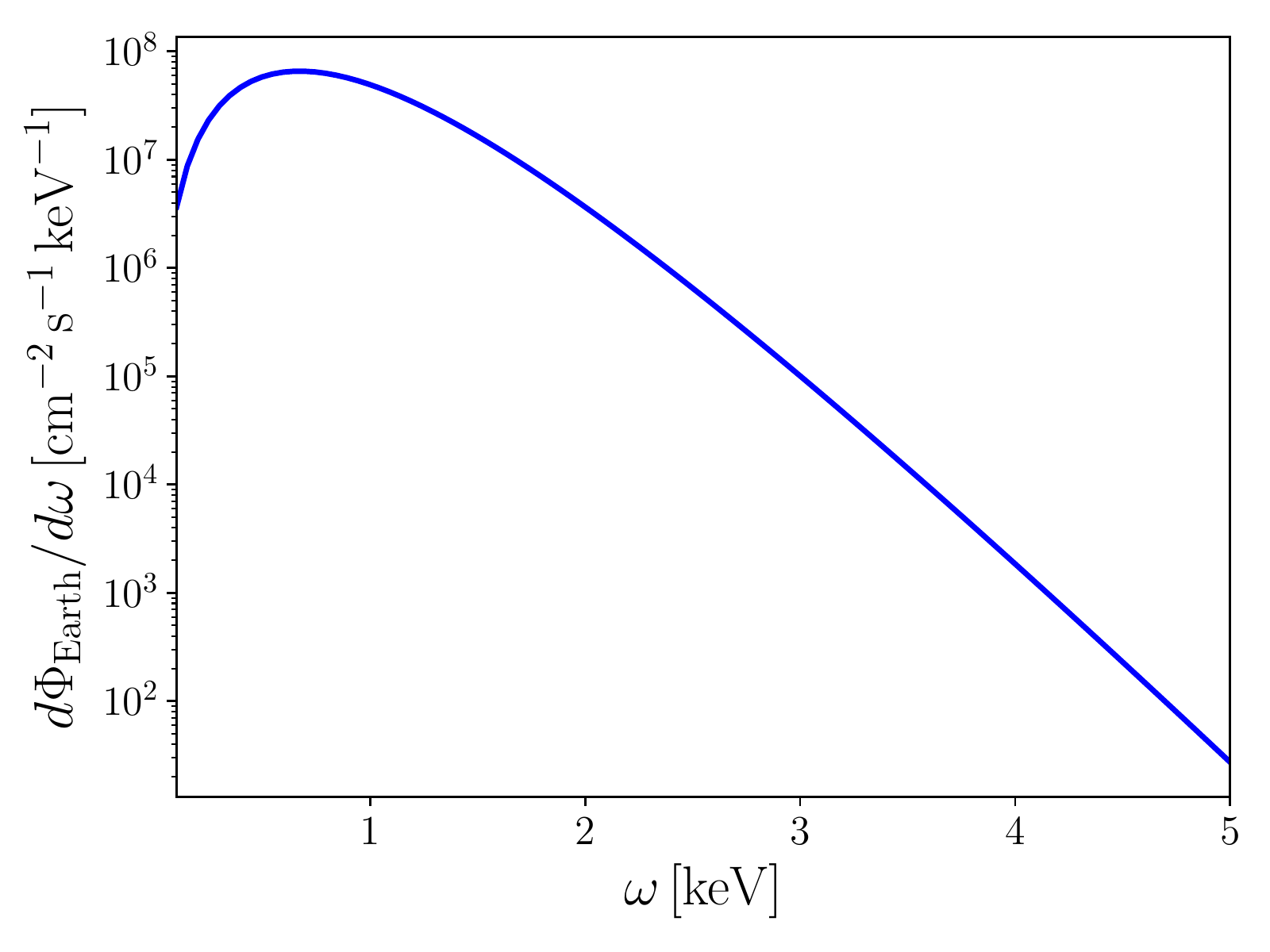}
    \caption{Solar chameleon differential flux per unit energy on Earth, in units of ${\rm cm}^{-2}\,{\rm s}^{-1}\,{\rm keV}^{-1}$, resulting from isotropic production in the solar tachocline. The chameleon parameters are fixed to $\beta_e=10^2$, $M_e=10^{3.6}\,{\rm keV}$, $\beta_{\gamma}=10^{10}$, $M_{\gamma}=1000\,{\rm TeV}$, $\Lambda=1\,\mu{\rm eV}$, and $n=1$. This is an example of set of parameters which maximises the resulting quality of fit to the XENON1T signal, as shown in Fig.~\ref{fig:fig1} (blue curve).}
    \label{fig:fig2}
\end{figure}

The detection rate in Eq.~(\ref{eq:diffrate_xenon}) depends on the spectrum of chameleons on Earth resulting from Primakoff-like production in the tachocline, as given by Eq.~(\ref{eq:Earthflux}). In Fig.~\ref{fig:fig2} we show this solar chameleon flux on Earth using the same benchmark choice of chameleon parameters as in Fig.~\ref{fig:fig1}. Note that the production flux remains high even at energies below $1\,{\rm keV}$, an aspect which will have important consequences for our subsequent discussion.

Although the fit in Fig.~\ref{fig:fig1} is visually adequate, some features require further investigation. In fact, focusing on the second and third bins, i.e.\ the two bins where the measured event rate deviates the most from $B_0$, the quality of the fit is only slightly worse that of the solar axion, neutrino magnetic moment, and tritium explanations invoked by XENON1T (see Figs.~7a-7d in Ref.~\cite{Aprile:2020tmw}). For comparison, the predicted signal resulting from the best-fit solar axion model is given by the red curve in Fig.~\ref{fig:fig1}. However, as discussed in Ref.~\cite{Aprile:2020tmw}, the background model is rejected at a significance of more than $3\sigma$ within all these signal models, much stronger than our $2.0\sigma$.

The paramount difference between the solar axion and solar chameleon models is the available production channels. The effects of this are evident by comparing the blue and red curves in Fig.~\ref{fig:fig1} (for solar chameleons and solar axions respectively). As discussed in the Introduction, screening prevents the production of chameleons in the core through bremsstrahlung, Primakoff, and Compton effects and proceeds via the Primakoff effect~\cite{Brax:2011wp} in the tachocline. Axions, on the other hand, are not affected by screening, and are produced in the core of the Sun through different mechanisms: these include ABC reactions~\cite{Raffelt:1985nk, Dimopoulos:1986kc, Redondo:2013wwa}, the Primakoff effect~\cite{Dicus:1978fp}, and the $^{57}$Fe transition line~\cite{Moriyama:1995bz}. These differences appear clearly in Fig.~\ref{fig:fig1}, where the solar chameleon model (blue line) shows one peak at $\omega \approx 2\,$keV, while the solar axion model (red line) shows three distinct peaks at three different energies corresponding to the different production mechanisms. These features result in a better fit of the solar axion model to the third, fourth, fifth, sixth, and 14$^{\rm th}$ bins, improving the overall fit and significance at which the model is preferred over the background. In particular, ABC reactions are responsible for the considerably improved fit to the bins from the third to the sixth. These differences between the solar chameleon and solar axion scenarios can be distinguished in future DM direct detection experiments by their different spectra. We further discuss the prospects for detection of solar chameleons in future DM direct detection experiments in Section~\ref{subsec:other}.

Although the measured recoil rate in the first bin is perfectly in line with the background model, the chameleon model overshoots this first point by just over one standard deviation. The reason is that the solar chameleon differential flux on Earth remains appreciable at low energies $\ll {\cal O}({\rm keV})$, see Fig.~\ref{fig:fig2}. This results in an increase in the event rate over $B_0$ at lower energies, leading to a poorer fit to the first bin, which is perfectly in line with $B_0$ and would not call for any additional contributions to the fit.

The reason why the signal beyond the third bin is completely dominated by the background model $B_0$ (which at that point is in very good agreement with the XENON1T measurements) is that the production flux, after peaking at an energy slightly below $\omega = 1\,{\rm keV}$, quickly drops for higher energies (see Fig.~\ref{fig:fig2}). In other words, in the sum in the numerator of Eq.~(\ref{eq:chi2}), one has $(\mathrm{d}R/\mathrm{d}\omega)_{\rm th} + B_0 \simeq B_0 \simeq (\mathrm{d}R/\mathrm{d}\omega)_{\rm meas}$ for $i \geq 4$, and therefore $(\mathrm{d}R/\mathrm{d}\omega)_{\rm th} + B_0 - (\mathrm{d}R/\mathrm{d}\omega)_{\rm meas} \ll 1$, resulting in small contributions to the $\chi^2$ for $i \geq 4$ in the numerator of Eq.~(\ref{eq:chi2}). For this reason, computing the $\chi^2$ over all 16 bins might be a misleading goodness-of-fit metric. For the same choice of benchmark parameters as mentioned above, and in Fig.~\ref{fig:fig1}, the contribution of the first three bins to the total $\chi^2$ is $\simeq 6.5$, which better quantifies the imperfect fit to the first three bins visible in Fig.~\ref{fig:fig1}.

\begin{figure}[!ht]
\includegraphics[width=1.0\linewidth]{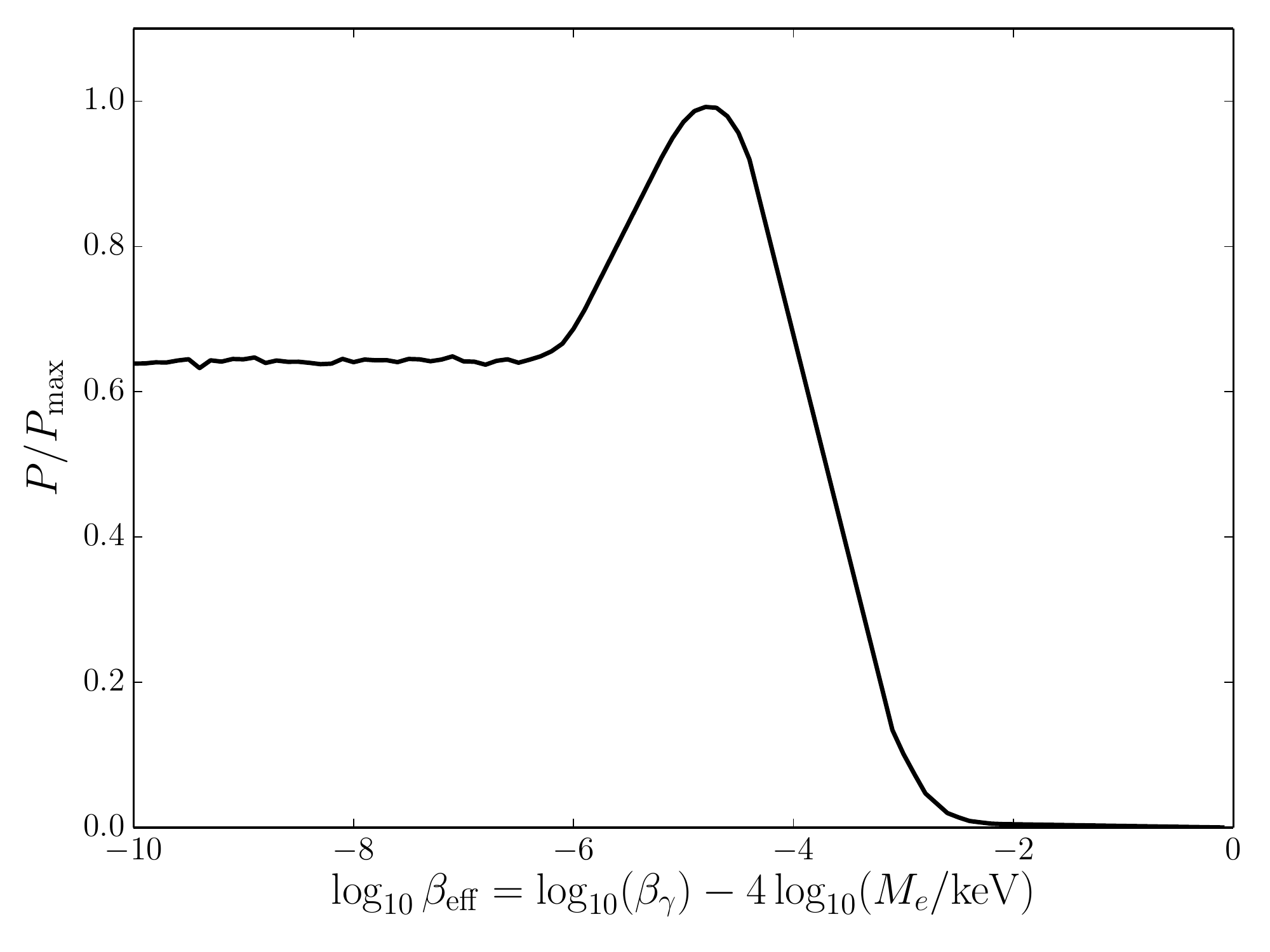}
\caption{Normalized posterior distribution for $\log_{10}\beta_{\rm eff}$, the combination of the photon coupling $\beta_{\gamma}$ and the electron disformal scale $M_e$ to which the XENON1T measurements are most sensitive, see Eqs.~(\ref{eq:betaeff},\ref{eq:betaeffbestfit}). A value of $\log_{10}\beta_{\rm eff} \simeq -4.5$ is required to provide a good fit to the XENON1T signal, of quality identical to that shown in Fig.~\ref{fig:fig1}.}
\label{fig:fig4}
\end{figure}

Finally, let us discuss how XENON1T constrains the effective coupling $\beta_{\rm eff}$, given by Eq.~(\ref{eq:betaeff}). We treat $\beta_{\rm eff}$ as a derived parameter whose posterior distribution we infer from our MCMC samples of the four fundamental parameters. The normalized posterior distribution for $\log_{10}\beta_{\rm eff}$ is given in Fig.~\ref{fig:fig4}, which shows that XENON1T is indeed able to set meaningful constraints on this parameter.

The shape of the $\log_{10}\beta_{\rm eff}$ posterior, with a tail as $\log_{10}\beta_{\rm eff} \to 0$ and a plateau for large negative values of $\log_{10}\beta_{\rm eff}$, is worth explaining further. Moving $\log_{10}\beta_{\rm eff} \to 0$ means that either $\beta_{\gamma}$ is being raised or $M_e$ is being lowered. In other words, either or both solar chameleon production and detection are being enhanced. Enhancing the signal sufficiently will make the total event rate too large compared to the XENON1T measurements, and hence increasingly unlikely, leading to the tail in the $\log_{10}\beta_{\rm eff}$ posterior as $\log_{10}\beta_{\rm eff} \to 0$.

On the other hand, moving $\log_{10}\beta_{\rm eff}$ towards large negative values means that either or both solar chameleon production and detection are being suppressed. Within this regime, the resulting event rate would be too low compared to the XENON1T detector background, and therefore the total signal is dominated by $B_0$. This implies that $(\mathrm{d}R/\mathrm{d}\omega)_{\rm th} + B_0 \simeq B_0$ for all $i$ in the numerator of Eq.~(\ref{eq:chi2}), regardless of the choice of model parameters. This behaviour leads to an extended ``plateau'' in parameter space where the likelihood is completely flat with $\chi^2 \simeq 17.2$. Along the plateau, the quality of the resulting fit to the XENON1T signal is not only identical for any choice of parameters, but also identical to the quality of a $B_0$-only fit. The goodness-of-fit along the plateau is worse by only $\Delta \chi^2=+4.0$ with respect to the solar chameleon best-fit (for which $\chi^2_{\min} \simeq 13.2$, see Fig.~\ref{fig:fig1}). This explains the shallow plateau in the $\log_{10}\beta_{\rm eff}$ posterior for large negative values of $\log_{10}\beta_{\rm eff}$ .

We find that the best fit to the XENON1T signal occurs for a value of $\beta_{\rm eff} \simeq 10^{-4.5}$. This implies
\begin{eqnarray}
\label{eq:betaeffbestfit}
\log_{10}\beta_{\gamma} - 4\log_{10} \left ( \frac{M_e}{{\rm keV}} \right ) \simeq -4.5\,.
\end{eqnarray}
Because of the shape of the $\log_{10}\beta_{\rm eff}$ posterior shown in Fig.~\ref{fig:fig4} (the tail can extend indefinitely to large negative values of $\log_{10}\beta_{\rm eff}$), we do not quote a confidence interval on $\log_{10}\beta_{\rm eff}$. Rather, we use Eq.~(\ref{eq:betaeffbestfit}) as indicative of what combinations of $\beta_{\gamma}$ and $M_e$ lead to a good fit to the XENON1T signal, of quality identical to that shown in Fig.~\ref{fig:fig1}. We have also verified our earlier expectation that $\beta_e$ and $M_{\gamma}$ play a negligible role in our analysis. We have fixed these parameters to $\beta_e=10^2$ and $M_{\gamma}=1000\,{\rm TeV}$ respectively, and repeated the analysis. Doing so, we find essentially the same posterior for $\log_{10}\beta_{\rm eff}$ as shown in Fig.~\ref{fig:fig4}, which was instead previously obtained by varying all four parameters.

Demanding that solar chameleons fit the XENON1T signal, and combining the relation in Eq.~(\ref{eq:betaeffbestfit}) and the upper limit of $\beta_{\gamma}<10^{11}$ from CAST~\cite{Cuendis:2019mgz} sets an upper limit on $M_e \lesssim 10\,{\rm MeV}$. In other words, if $M_e \gtrsim 10\,{\rm MeV}$, for any allowed value of $\beta_{\gamma}$ the event rate will be completely dominated by $B_0$, and we will find ourselves along the plateau for large negative values of $\log_{10}\beta_{\rm eff}$ in Fig.~\ref{fig:fig4}. This upper limit on $M_e$ is ostensibly in contradiction with the lower limit one obtains by demanding that horizontal branch stars are not affected by the disformal coupling, which sets $M_e \gtrsim 0.1\,{\rm GeV}$, as found by one of us~\cite{Brax:2014vva}. However, we note that the limit obtained in Ref.~\cite{Brax:2014vva} is not applicable to our case, as it was obtained assuming that the scalar is massless. This is clearly not the case in our scenario, given the constraints we have imposed on the chameleon mass within the Sun, which in turn suppresses production, making it so that the bounds obtained in Ref.~\cite{Brax:2014vva} do not apply. This is generically true for all of the relevant limits in Ref.~\cite{Brax:2014vva}, which were all derived assuming a massless scalar.

The upper limit $M_e \lesssim 10\,{\rm MeV}$ is nominally in tension with LEP constraints, which require $M_e \gtrsim 3\,{\rm GeV}$ as derived by one of us in Ref.~\cite{Brax:2014vva}. However, care must be taken with this bound, as it again was derived assuming a massless chameleon. A proper re-evaluation of the bounds of Ref.~\cite{Brax:2014vva} would require a dedicated analysis, for instance determining the field profile in the LEP pipe simultaneously including conformal and disformal couplings, a calculation which is well beyond the scope of this paper. Therefore, in continuing our exciting program for the direct detection of dark energy quanta, re-evaluating the LEP bounds is of paramount importance, and will be the subject of follow-up work.~\footnote{Should the model be in tension with LEP constraints, extensions which can alleviate this tension are possible. For instance, along the lines of Ref.~\cite{Brax:2012hm}, one could entertain the possibility of an environmentally-dependent disformal coupling. This effectively amounts to an extension of the chameleon mechanism retaining the background field-dependence in the disformal term in Eq.~(\ref{eq:jordaneinstein}), which we dropped for simplicity.}

Let us summarize the main findings of this Section:
\begin{enumerate}
\item The expected event rate in the XENON1T detector is sensitive to the combination of the photon coupling $\beta_{\gamma}$ and the electron disformal scale $M_e$ given by the effective coupling $\beta_{\rm eff}$ in Eq.~(\ref{eq:betaeff}).
\item On the other hand, XENON1T has only very weak sensitivity to $\beta_e$, $M_{\gamma}$, $\Lambda$, and $n$.
\item A value of $\log_{10}\beta_{\rm eff} \simeq -4.5$ is required to fit the XENON1T signal well. This leads to an improvement in $\chi^2$ of $\simeq 4.0$ compared to a $B_0$-only fit ($B_0$ excluded at $2.0\sigma$), and a quality of fit as shown in Fig.~\ref{fig:fig1}. In no region of parameter space can the quality of the fit to the XENON1T signal be better than that shown in Fig.~\ref{fig:fig1}. As CAST requires $\beta_{\gamma}<10^{11}$, demanding that solar chameleons explain the XENON1T signal and therefore $\log_{10}\beta_{\rm eff} \simeq -4.5$ implies $M_e \lesssim 10\,{\rm MeV}$.
\item Given the previous points 1.~and 2., we can identify various combinations of the chameleon parameters which lead to a fit of identical quality to that shown in Fig.~\ref{fig:fig1}. There is therefore a large window of parameter space which can account for part of the XENON1T excess, while remaining consistent with laboratory and astrophysical tests.
\item With respect to the solar axion model invoked by XENON1T, the statistical significance of the preference for the solar chameleon model is lower because of the poorer fits to the first bin, as well as to a few bins at higher energies (due to the larger available number of production channels for solar axions). However, we stress that the solar chameleon model is not excluded by other bounds, unlike the solar axion model.
\end{enumerate}
Our overall conclusion is that solar chameleons are able to provide an adequate fit to the XENON1T signal. This raises the tantalizing possibility that XENON1T, originally constructed to detect dark matter, may have achieved the first direct detection of dark energy quanta.

In principle, one may also consider a hybrid chameleon-axion scenario where both particles are present and contribute to the XENON1T signal, which thus results from an incoherent sum of solar-produced axions and chameleons. This could be beneficial for the solar axion model, as it might be able to alleviate the tension in the axion-electron coupling between the XENON1T results and astrophysical constraints~\cite{DiLuzio:2020jjp}. Within this scenario, the lower-energy excess would be fitted by solar chameleons, plus a smaller contribution from solar axions, allowing for a lower value of $g_{ae}$. On the other hand, the higher-energy end would be fit by the solar axion via its couplings to photons and nucleons. We defer the study of this interesting hybrid possibility to follow-up work.

\section{Discussion}
\label{sec:discussion}

In this section, we  discuss other experimental bounds on our scenario, stellar bounds in particular, and the prospects for detecting dark energy in current and planned dark matter direct detection experiments.

\subsection{Stellar cooling constraints}
\label{sec:stellarconstraints}

The stellar bounds on the axion-electron and axion-photon coupling are debilitating for the solar axion interpretation of the XENON1T excess~\cite{DiLuzio:2020jjp,Vinyoles:2015aba}. The situation with chameleons is different. The paramount difference between the two models is the environment-dependence of the chameleon's mass. This ensures that chameleons are not produced in the Sun's core since Compton and bremsstrahlung processes are kinematically suppressed, leading to a severe Boltzmann suppression. Instead, chameleons are produced in the strong magnetic field of the solar tachocline. Similarly, the cores of red giant, horizontal branch (HB), and white dwarf stars are denser than the Sun's by a significant factor, implying an even stronger suppression in these objects. It is possible that some of these objects may have strong magnetic fields~\cite{2018MNRAS.478..899B}, but without dedicated individual observations and detailed stellar modelling, it is not possible to derive quantitative constraints on chameleons.

Additionally, in Tab.~\ref{tab:cooling} we report typical core densities and temperatures for these objects, alongside the chameleon mass for the benchmark parameter space point we have considered throughout the paper. As we see, production of chameleons within these objects is strongly kinematically suppressed, even more so than within the Sun, implying that stellar cooling constraints are evaded. Therefore, the bounds obtained by one of us in Ref.~\cite{Brax:2014vva}, derived assuming a massless chameleon, may be safely evaded.

\begin{table}[!ht]
\def\arraystretch{1.5}
	\begin{tabular}{cccc}
    \cline{2-3}
    \hline\hline
		Stellar object & $\rho_{\rm core}$ & $T_{\rm core}$ & $m_{\rm core}$ \\
		& (typical) & (typical) & \\
		& [${\rm g}/{\rm cm}^3$] & [${\rm keV}$] & [${\rm keV}$]  \\
		\hline
		Sun & $150$ & $1.3$ & $6$ \\
		White dwarfs & $10^6$ & ${\cal O}(1)$ & $\sim 6000$ \\
		Red giants & $5 \times 10^5$ & ${\cal O}(10)$ & $\sim 4000$ \\
		Horizontal branch stars & $5 \times 10^4$ & ${\cal O}(10)$ & $\sim 100$ \\
		\hline
		\hline
	\end{tabular}
	\caption{Typical core densities and temperatures for stellar objects of interest: the Sun, white dwarfs, red giants, and horizontal branch stars. The final column reports the chameleon mass within the core of these objects for the benchmark parameter space point we have considered throughout the paper, where $\beta_e=10^2$, $\Lambda=1\,\mu{\rm eV}$, and $n=1$. It is clear that production of stellar chameleons is strongly kinematically suppressed within these objects, as $m_{\rm core} \gg T_{\rm core}$.}
	\label{tab:cooling}
\end{table}

Finally, we note that the literature is rich with \emph{astrophysical} (stellar and galactic) bounds on chameleons, see e.g.\ Refs.~\cite{Davis:2011qf, Jain:2012tn, Sakstein:2018fwz, Baker:2019gxo,Naik:2018mtx,Naik:2019moz, Naik:2020uby, Desmond:2020gzn}. These refer to searches for the effects of fifth forces rather than chameleon particle production. Chameleon models predicting fifth forces in astrophysical objects occupy a different region of parameter space than those that give rise to chameleon particle production in the solar tachocline considered in this work. The underlying reason for this is that fifth forces are only relevant in astrophysical bodies of radius $R$ if $m(\varphi_0)R\sim 1$, where $\varphi_0$ is the background field in that body. The chameleon theories accessible to the direct detection experiments that we have discussed in this work have $m(\varphi_\odot)R_\odot\gg1$, implying a fifth force range too small to affect stellar structure. For this reason, it is generally the case that astrophysical fifth force searches do not constrain our proposed scenario. These theories may be subject to the bounds arising from laboratory tests~\cite{Burrage:2016bwy,Burrage:2017qrf}, although such bounds are highly model-dependent. The specific model studied in this work is able to simultaneously satisfy all experimental bounds and account for part of the XENON1T signal.

\subsection{Other dark matter direct detection experiments}
\label{subsec:other}

\begin{table}[!ht]
\def\arraystretch{1.5}
	\begin{tabular}{cccc}
    \cline{2-3}
    \hline\hline
		Experiment & Exposure & Electron recoil & Events / yr\\
		& & background & (expected)\\
		& (${\rm ton}\!\times\!{\rm yr}$) & (${\rm ton}\!\times\!{\rm yr}\!\times\!{\rm keV}$)$^{-1}$ & \\
		\hline
		XENON1T~\cite{Aprile:2020tmw} & 0.65 & 76.0 & 20\\
		XENONnT~\cite{Aprile:2020vtw} & 20.0 & 12.3 & 180\\
		PandaX-4T~\cite{Zhang:2018xdp} & 5.6 & 18.0 & 130\\
		LUX-ZEPLIN~\cite{Akerib:2018lyp} & 15.0 & 14.0 & 250\\
		\hline
		\hline
	\end{tabular}
	\caption{Expected exposure in units of ${\rm ton}\times{\rm yr}$ (tonne-year) and electron recoil background in units of (${\rm ton}\times{\rm yr}\times{\rm keV}$)$^{-1}$ for recoil energies $\lesssim 10\,{\rm keV}$, expected for the upcoming XENONnT~\cite{Aprile:2020vtw}, PandaX-4T~\cite{Zhang:2018xdp}, and LUX-ZEPLIN~\cite{Akerib:2018lyp} experiments, which will be able to confirm or disprove the XENON1T excess. The last column reports the number of excess events that are expected per year in each detector, in the energy range $(1-30)\,$keV.}
	\label{tabParams}
\end{table}

Recent DM direct detection searches prior to XENON1T did not report any excess over the expected background. For example, the PandaX-II experiment with an exposure of $\approx27\,{\rm ton}\times{\rm day}$ placed an upper limit on the axion-electron coupling $g_{ae} \lesssim 4.35\times 10^{-12}$ for an axion mass $m_a \lesssim 1\,$keV~\cite{Fu:2017lfc}.\footnote{A more recent analysis with an exposure of $100.7\,{\rm ton}\times{\rm day}$ found a similar result, $g_{ae} \lesssim 4.6\times 10^{-12}$~\cite{Zhou:2020bvf}.} A competitive limit has also been placed by the LUX-ZEPLIN collaboration with an exposure of $11.2\,{\rm ton}\times{\rm day}$, which lead to the result $g_{ae} \lesssim 3.5 \times 10^{-12}$ at 90\% confidence level~\cite{Akerib:2017uem}. These exposures are all significantly lower than the XENON1T exposure of $0.65\,{\rm ton}\times{\rm yr}$, which could explain why these experiments did not observe any low-energy excess.

Various upcoming experiments plan to search for the signal reported by the XENON1T collaboration, and will either confirm or disprove it. These experiments include XENONnT (the planned upgrade to XENON1T)~\cite{Aprile:2020vtw}, as well as PandaX-4T~\cite{Zhang:2018xdp} and LUX-ZEPLIN~\cite{Akerib:2018lyp}, all of which use a dual-phase xenon time projection chamber. In Table~\ref{tabParams} we report the expected exposure in units of ${\rm ton}\times{\rm yr}$ and electron recoil background in units of (${\rm ton}\times{\rm yr}\times{\rm keV}$)$^{-1}$ for each of these experiments.

We focus on the benchmark point in parameter space considered in Sec.~\ref{sec:results}, as well as Figs.~\ref{fig:fig1} and~\ref{fig:fig2}. Adopting these values, the expected excess number of events per year due to a hypothetical signal is about 20 for XENON1T, 180 for XENONnT, 130 for PandaX-4T, and 250 for LUX-ZEPLIN. Note that we have not considered the effects of background noise or energy resolution in obtaining these estimates of excess number of events since the energy resolution specifications for these future experiments are currently unavailable.

As is clear from Table~\ref{tabParams}, all of these next-generation xenon-based experiments will have a background level $B_0$ a factor of $\approx 5$-$6$ lower than current levels, while the exposure will increase by over an order of magnitude. This combination of lower background and increased exposure results leads to the extremely high number of expected events. By virtue of this, future experiments will be able to confirm or disprove our hypothesis that solar chameleons are the origin of the XENON1T signal with extremely high statistical significance. More generally, it will be possible to test at high significance whether the XENON1T excess is due to a statistical fluke, a background contaminant, or new physics such as the scenario considered here.

\section{Conclusions}
\label{sec:conclusions}

Most of our knowledge about dark energy (DE) arises from cosmological measurements which are mainly sensitive to its gravitational effects. Yet searching for non-gravitational signatures of DE by directly detecting DE quanta would be an extremely important step towards understanding the physics powering cosmic acceleration. In this paper, our aim has been to broaden the scope of new physics accessible to terrestrial dark matter (DM) direct detection experiments by investigating the intriguing possibility that these instruments may be able to detect DE quanta via their couplings to matter. Specifically, we have envisaged a scenario wherein DE particles produced in the strong magnetic fields of the solar tachocline travel to Earth and are absorbed by electrons or nuclei in terrestrial DM detectors. In this paper, focusing on DE scalars including screening mechanisms of the chameleon type, we have laid out the formalism for computing the expected signal from such a process, demonstrating that it can lead to measurable recoils in the ${\cal O}({\rm keV})$ range, well within the sensitivity of current and upcoming DM direct detection experiments.

We have applied our results to the XENON1T experiment, which recently reported a $\approx 3.3\sigma$ excess in their electron recoil data at recoil energies of $\approx 1$-$2\,{\rm keV}$. We have shown that solar chameleons can explain the XENON1T excess (see blue curve in Fig.~\ref{fig:fig1}), and are preferred over the background-only hypothesis at a significance of $\approx 2.0\sigma$. Our results have been obtained using the code which we make publicly available at \href{https://github.com/lucavisinelli/XENONCHAM}{github.com/lucavisinelli/XENONCHAM}.

Compared to the much discussed solar axion interpretation of the XENON1T signal, the statistical preference for solar chameleons is lower, mostly due to the reduced number of available production channels in the Sun (chameleons are only produced through Primakoff-like processes in the tachocline). However, the stellar cooling constraints which are debilitating for the solar axion model do not apply to solar chameleons due to the environment-dependence of the chameleon mass, which within the dense environments of red giants and white dwarfs results in Compton and bremsstrahlung production processes being suppressed.

We have also studied prospects for testing this explanation in future DM direct detection experiments. If solar chameleons are indeed at the origin of the XENON1T excess then this will be confirmed at very high significance in upcoming experiments such as XENONnT, PandaX-4T, and LUX-ZEPLIN. What is perhaps more important is that future low-threshold DM direct detection experiments will be well-suited to detect the signatures of DE particles produced within the Sun.

There are several avenues for future research in this direction. While our study has focused on chameleon-screened scalars, we stress that the effective theory approach we have adopted is quite general, and our formalism can therefore be applied (with appropriate modifications) to several other physical scenarios of interest. More generally, we hope that this paper will stimulate further research aimed towards enabling direct detection of dark energy, searching for non-gravitational signatures of dark energy, and unraveling the physics of cosmic acceleration in terrestrial laboratories.

\begin{acknowledgments}
\noindent We are grateful to Djuna Croon, Samuel D. McDermott, Rouven Essig, Katherine Freese, Chris Kelso, Jason Kumar, Xudong Sun, and Sebastian Trojanowski for several enlightening discussions. S.V.\ acknowledges support from the Isaac Newton Trust and the Kavli Foundation through a Newton-Kavli Fellowship, and by a grant from the Foundation Blanceflor Boncompagni Ludovisi, n\'{e}e Bildt. S.V. acknowledges a College Research Associateship at Homerton College, University of Cambridge. L.V.\ acknowledges support from the European Union's Horizon 2020 research and innovation programme under the Marie Sk{\l}odowska-Curie grant agreement ``TALeNT'' No.~754496 (H2020-MSCA-COFUND-2016 FELLINI), as well as support from the NWO Physics Vrij Programme ``The Hidden Universe of Weakly Interacting Particles'' with project number 680.92.18.03 (NWO Vrije Programma), which is (partly) financed by the Dutch Research Council (NWO). P.B.\ acknowledges support from the Institut Pascal at Universit{\'e} Paris-Saclay with the support of the P2I and SPU research departments and the P2IO Laboratory of Excellence (program ``Investissements d'avenir'' ANR-11-IDEX-0003-01 Paris-Saclay and ANR-10-LABX-0038), as well as by the IPhT. A.C.D.\ acknowledges partial support from the STFC Consolidated Grants No.~ST/P000673/1, No.~ST/P000681/1, and No.~ST/T000694/1. This work was performed using resources provided by the Cambridge Service for Data Driven Discovery (CSD3) operated by the University of Cambridge Research Computing Service (\href{https://www.hpc.cam.ac.uk/}{www.hpc.cam.ac.uk}), provided by Dell EMC and Intel using Tier-2 funding from the Engineering and Physical Sciences Research Council (capital grant EP/P020259/1), and DiRAC funding from the Science and Technology Facilities Council (\href{https://www.dirac.ac.uk/}{www.dirac.ac.uk}).
\end{acknowledgments}

\appendix

\section{Chameleon Screening}
\label{sec:AppendixA}

Chameleon-screened theories provide an explicit example  where {the mapping between the effective theory in the solar system given in the main text}  and a more complete theory can be calculated~\cite{Brax:2011aw} (see Refs.~\cite{Sakstein:2015oqa,Burrage:2017qrf} for more details about screening). In this case, the scalar $\phi$ is canonically normalised with a scalar potential $V(\phi)$. The interaction with matter is obtained via the Jordan frame metric $g^J_{\mu\nu}= A^2(\phi) g_{\mu\nu}$. We treat the disformal and derivative interactions as negligible perturbations compared to the dominant effect due to the conformal rescaling given by $A(\phi)$. This is valid provided the suppression scales of the disformal and derivative interactions are large enough. In the Einstein frame, where the graviton and scalar are canonically normalized but the coupling to matter is non-minimal, the dynamics of $\phi$ depend on the effective potential
\begin{equation}
\label{eq:effectivepotential}
V_{\rm eff} (\phi)= V(\phi) + \rho A(\phi)\,.
\end{equation}
where $\rho$ is the conserved matter density in the Einstein frame, related to the density in the Einstein frame as $\rho_E= A \rho$. This relationship between $\rho_E$ and $\rho$ follows from the non-conservation of the Einstein-frame energy momentum tensor which, for pressureless matter, gives~\cite{Brax:2017idh}
\begin{equation}
\dot \rho_E +3h_E \rho_E= \frac{\beta}{m_{\rm Pl}} \rho_E \dot \phi\,,
\label{rE}
\end{equation}
where $\dot \rho_E= u^\mu D_\mu \rho_E$ is the Einstein-time derivative along the trajectories of the matter particles with 4-velocities $u^\mu$. The local Hubble rate in the Einstein frame is  $h_E= D_\mu u^\mu/3 $  which reduces to the cosmological Hubble rate $h_E=H$ on large scales. From Eq.~\eqref{rE}, we deduce that
\begin{equation}
\dot \rho +3h_E \rho=0\,,
\label{rE1}
\end{equation}
which expresses the local conservation of the matter density $\rho$. The Jordan frame matter density $\rho_J=A^{-4} \rho_E$ is conserved in the Jordan frame where $\mathrm{d}t_J= A \mathrm{d}t_E$ and the local Hubble rate is given by $h_J= \frac{h_E}{A} + \frac{\dot A}{A^2}$, i.e.\
\begin{equation}
    \frac{d\rho_J}{dt_J}+ 3 h_J \rho_J=0\,.
\end{equation}
The Jordan frame matter density is deemed to be the ``physical" density as, in the local Jordan frame where the Jordan metric is nearly Minkowskian, the Lagrangian of the standard model reduces to the usual one. As long as $A\approx 1$, the difference between $\rho$ and $\rho_J$ is negligible. 

Chameleon-screened models have a minimum of the effective potential $\phi(\rho)$ which depends on the conserved matter density $\rho$ in the Einstein frame. The minimum equation reads
\begin{equation}
\left.\frac{\dd V}{\dd\phi}\right\vert_{\phi(\rho)}+ \rho \left.\frac{\dd A}{\dd\phi}\right\vert_{\phi(\rho)}=0\,,
\end{equation}
which can be used to obtained a parametric description of the value $\phi (\rho)$. Taking the derivative of the minimum equation with respect to $\rho$ leads to
\begin{equation}
\label{min}
m_\phi^2(\rho) \frac{\dd\phi(\rho)}{\dd\rho} = - \left.\frac{\dd A}{\dd\phi}\right\vert_{\phi(\rho)}\,,
\end{equation}
where we have defined the effective mass
\begin{equation}
    \label{eq:def_effectivemass}
    m_\phi^2 (\rho)= \left.\frac{{\mathrm d}^2V}{{\mathrm d}\phi^2}\right\vert_{\phi(\rho)}+ \rho \left.\frac{{\mathrm d}^2A}{{\mathrm d}\phi^2}\right\vert_{\phi(\rho)}\,.
\end{equation}
It is convenient to introduce the effective coupling
\begin{equation}
    \label{eq:definecoupling}
    \beta (\rho)= \mpl\left.\frac{\dd\ln A}{\dd\phi}\right\vert_{\phi(\rho)}\,,
\end{equation}
and to integrate (\ref{min})
\begin{equation}
\frac{\phi (\rho)}{\mpl} = \phi_0 - \int_{\rho_0}^\rho \frac{ A(\rho) \beta (\rho)}{m_\phi^2(\rho)\mpl^2} \dd\rho\,.
\end{equation}
This provides a one-to-one relationship between the density of matter and the value of the minimum $\phi(\rho)$. 
Moreover from (\ref{min}) we get
\begin{equation}
    \label{varA}
    \frac{\dd A^{-1}(\rho)}{\dd\rho}= \frac{\beta^2 (\rho)}{m_\phi^2(\rho) \mpl^2}\,,
\end{equation}
allowing one to obtain $A(\rho)$ as a decreasing function of $\rho$, i.e.\ $A^{-1}$ is an increasing function of $\rho$ with a positive derivative. Finally the minimum equation gives
\begin{equation}
    \frac{\dd V}{\dd\rho}= - \rho \frac{\beta^2(\rho) A^2(\rho)}{m_\phi^2(\rho) \mpl^2}\,,
\end{equation}
from which we can find $V(\rho)$. Hence eliminating $\rho$ between $A(\rho)$, $V(\rho)$, and $\phi(\rho)$, one can reconstruct $A(\phi)$ and $V(\phi)$.

The potential $V(\phi)$ is defined up to an integration constant, i.e.\ the screening properties of the models do not depend on an additive cosmological constant. This constant has to be tuned to generate the appropriate acceleration of the Universe. As the screening properties are independent of this choice, this has no effects on the results obtained in this paper.
As a specific example, let us consider the inverse power law potential. 
\begin{equation}
    \label{eq:potential}
    V(\phi) = V_0+ \frac{\Lambda^{4+n}}{\phi^n}\,,
\end{equation}
where we have included the constant $V_0$ which needs to be adjusted to fit the current dark energy value. As we have seen, the screening properties only depend on the the inverse power law part. We note that models could arise from the strong dynamics of a confining supersymmetric dark sector at higher energy \cite{Binetruy:1998rz}.

The configuration of the field at which the potential is minimized $\phi (\rho)$, and the chameleon rest mass squared obtained from the curvature of the effective potential, $m_\phi^2$, are given by
\begin{eqnarray}
    \phi(\rho) &=& \left(\frac{n \, \mpl\, \Lambda^{4 + n}}{\beta_m\, \rho}\right)^{\frac{1}{1 + n}}\,,\label{eq:phimin}\\
    m_\phi^2(\rho) &=& n( 1+n)\Lambda^{4+n}\left(\frac{\beta_m\,\rho}{n\mpl\Lambda^{4+n}}\right)^{\frac{2+n}{1+n}}\,,
\end{eqnarray}
This is the mass $m_\phi$ that we use as a template in the main text [see Eq.~\eqref{eq:mphi_main}].

Finally this reconstruction procedure of $V(\phi)$ and $A(\phi)$ from $m_\phi(\rho)$ and $\beta(\rho)$ allows one to design models where the production of scalars in the tachocline is favoured compared to very deep inside the Sun or in other even denser stars. In the Sun, all that is required is that in the core  where the density $\rho_{\rm core} \approx 150{\rm \,g\,cm^{-3}}$ is large compared to the one in the tachocline $\rho_{\rm tach}\approx 1{\rm \,g\,cm^{-3}}$, the production of scalars is kinematically forbidden. For chameleons produced in matter, e.g.\ by the Primakoff process deep in the electric field of a nucleus, their effective mass is modified by the presence of the surrounding plasma and their mixing with  photons as  $m_{\rm eff}^2= m_\phi^2(\rho_{\rm core})- \omega^2_{\rm Pl}$ where the plasma frequency is defined in Eq.~\eqref{plas}. For large enough densities, the mass  $m_\phi(\rho_{\rm core})$ is generically larger than the plasma frequency which scales as $\rho^{1/2}$.\footnote{For the inverse power law chameleons with $n=1$, their mass scales as $\rho^{3/4}$.} As a result, only scalars of momenta $k^2\simeq T^2_{\rm core}-m^2_{\rm eff} $ can be produced. When $m_{\rm eff}\gtrsim T_{\rm core}$,  production is highly suppressed.  This also applies to the Compton and bremsstrahlung processes involving the direct coupling between scalars and electrons where the same kinematical obstruction is at play. This mechanism was first proposed in Ref.~\cite{Brax:2007ak}.

\section{Production of chameleons in the Sun}

\label{sec:production}

Given the chameleon-photon disformal coupling in Eq.~\eqref{eq:LagrangianEM}, the probability of a photon in the uniform magnetic field $B$ converting into a chameleon after a distance $\ell$ is given by~\cite{Brax:2015fya}
\begin{equation}
    \label{eq:probcham}
    P_{\gamma\to\phi} = \frac{4\Delta_B^2}{4\Delta_B^2 + (\Delta_{\rm pl}-\Delta_a)^2}\,\sin^2\frac{\ell}{\ell_\omega}\,,
\end{equation}
where the coefficients are
\begin{eqnarray}
    \Delta_B &=& \frac{2\beta_\gamma}{\mpl}\frac{B}{\sqrt{1+b^2}}\,,\\
    \Delta_{\rm pl} &=& \frac{\omega_{\rm pl}^2}{2\omega}\,,\\
    \Delta_a &=& \frac{m_\phi^2 + 2b^2\omega^2\left(1-B_z^2/B^2\right)}{2\omega\left(1+b^2\right)}\,,\label{eq:Deltaa}\\
    \ell_\omega &=& \frac{2}{\sqrt{4\Delta_B^2 + (\Delta_{\rm pl}-\Delta_a)^2}}\,.\label{eq:ellomega}
\end{eqnarray}
In the expressions above, $\omega$ is the energy of the produced chameleon, the dimensionless parameter $b = B_t^2/M_\gamma^4$ is the ratio of the magnetic field in the solar tachocline $B_t$ to the UV-cutoff scale of the effective theory, and the plasma frequency is given in terms of the electron number density $n_e$ and the electron mass $m_e$ as
\begin{equation}
	\omega_{\rm pl}^2 = \frac{4\pi\,n_e}{m_e} \approx (2.0\times 10^8{\rm \,GHz})^2\,\left(\frac{n_e}{10^{23}{\rm \,cm^{-3}}}\right)\,.
	\label{plas}
\end{equation}
The quantity $B_z$ is the $z$-component of the magnetic field which we fix by assuming an isotropic magnetic field distribution as $B_z^2 = B^2/3$.

The thickness of the tachocline is much larger than the main free path of photons in the region, $\lambda \approx 0.3\,$cm~\cite{Blancard_2011, Krief:2016znd}, so that photon propagation in this region proceeds through a random walk process, which can be described as a Poisson diffusion process with mean free path $\lambda$. For a typical distance $\ell$ between two scatterings, the total number of scatterings per unit time is $\sim c/\ell$. For a given length path $\ell$, the differential probability of conversion in the solar interior is~\cite{Brax:2011wp}
\begin{equation}
    \label{eq:conversionprob}
    \frac{\mathrm{d}P_\phi}{\mathrm{d}R} = \int_0^{+\infty}\frac{\mathrm{d}\ell}{\ell}\sqrt{\frac{{\rm ls}}{\ell}}\,\frac{e^{-\ell/\lambda}}{\lambda}\,P_{\gamma\to\phi}\,,
\end{equation}
where ${\rm ls} = {c \bar t}\simeq 3\times 10^{10}\,$cm in the tachocline, i.e.\ approximately one light-second. Here, $\bar t$ is the typical time such that the photon flux at the tachocline $n_{\gamma, t}= \bar v \bar n_t$ where $\bar n_t$ is the photon number density at the tachocline and $\bar v = (c\lambda/\bar t)^{1/2}$ is the typical radial velocity of photons due to their Brownian motion. The differential flux of chameleons per unit energy emitted by the Sun is
\begin{equation}
    \label{eq:champroduction}
	\frac{\mathrm{d}\Phi}{\mathrm{d}\omega} = \int_0^{R_\odot}\mathrm{d}R\,p_\gamma(R)\,n_\gamma(R)\,\frac{\mathrm{d}P_\phi}{\mathrm{d}R}\,,
\end{equation}
where $n_\gamma(R)$ is the photon flux profile and the photon spectrum $p_\gamma(R)$ depends on the temperature profile of the plasma $T = T(R)$ as
\begin{equation}
    \label{eq:photonspectrum}
	p_\gamma(R) = \frac{1}{2\zeta(3)\,T^3}\,\frac{\omega^2}{\exp(\omega/T)-1}\,.
\end{equation}

We model the magnetic field profile inside the Sun as a thin shell around the solar tachocline, where the magnetic field is taken to be constant with a value $B_t = 30\,$T. The thin shell around the tachocline has radius $R_t = 0.7R_\odot$, where $R_\odot$ is the radius of the Sun, and the thickness $\Delta R = 0.01R_\odot$. The integrand in Eq.~\eqref{eq:champroduction} at the tachocline is then
\begin{equation}
    \label{eq:champroduction_t}
	\frac{\mathrm{d}\Phi}{\mathrm{d}\omega} = \Delta R\,p_{\gamma, t}\,n_{\gamma, t}\,\int_0^{+\infty}\frac{\mathrm{d}\ell}{\ell}\sqrt{\frac{{\rm ls}}{\ell}}\,\frac{e^{-\ell/\lambda}}{\lambda}\,P_{\gamma\to\phi}\,,
\end{equation}
where $n_{\gamma, t} = n_\gamma(R_t) \approx 10^{21}\,{\rm cm}^{-2}\,{\rm s}^{-1}$ and $p_{\gamma, t}$ is the expression in Eq.~\eqref{eq:photonspectrum} evaluated at the tachocline temperature $T \approx 0.2\,{\rm keV}$, and we have used Eq.~\eqref{eq:conversionprob} which expresses the differential probability of conversion. Inserting Eq.~\eqref{eq:probcham} into Eq.~\eqref{eq:champroduction_t} we obtain
\begin{equation}
    \label{eq:production}
	\frac{\mathrm{d}\Phi}{\mathrm{d}\omega} = p_{\gamma, t}\,n_{\gamma, t}\,\frac{\Delta R}{\lambda} \,\frac{4\Delta_B^2}{4\Delta_B^2 + (\Delta_{\rm pl}-\Delta_a)^2}\,\sqrt{\frac{\rm ls}{\ell_\omega}}\,\mathcal{I}\left(\frac{\ell_\omega}{\lambda}\right)\,,
\end{equation}
where the integral over $y = \ell/\ell_\omega$ has been performed exactly. For any constant $a$, we find
\begin{equation}
    \label{eq:integral}
    \mathcal{I}(a) \!\equiv\! \int_0^{+\infty}\mathrm{d}y \frac{\sin^2y}{y^{3/2}}e^{-a y} \!=\! \sqrt{\frac{\pi}{2}}\left(\!\sqrt{a \!+\! \sqrt{a^2 + 4}} \!-\! \sqrt{2 a}\!\right) \,.
\end{equation}
For most of the region of the parameter space we explore, the relation $\ell_\omega \ll \lambda$ holds, for which the integral in Eq.~(\ref{eq:integral}) is $\mathcal{I}(\ell_\omega/\lambda) \approx \sqrt{\pi}$.

For the region of parameters allowed we have $\Delta_B \ll \Delta_{\rm pl}$. We also assume $M_\gamma \gg {\cal O}({\rm keV})$, which corresponds to $b \ll 1$. In this limit, the expression for the solar chameleon flux in Eq.~\eqref{eq:production} reduces to
\begin{equation}
    \label{eq:production1}
	\frac{\mathrm{d}\Phi}{\mathrm{d}\omega} = p_{\gamma, t}\,n_{\gamma, t}\,\frac{\Delta R}{\lambda} \,\frac{32\beta_\gamma^2\,{\rm ls}^{1/2} B^2}{\mpl^2}\left(\frac{\omega}{\omega_{\rm pl}^2 - m_\phi^2}\right)^{3/2}\,\mathcal{I}\left(\frac{\ell_\omega}{\lambda}\right)\,.
\end{equation}
In the limit where $m_\phi^2 \ll \omega_{\rm pl}^2$, Eq.~\eqref{eq:production1} reduces to
\begin{equation}
    \label{eq:production2}
	\frac{\mathrm{d}\Phi}{\mathrm{d}\omega} = p_{\gamma, t}\,n_{\gamma, t}\,\frac{\Delta R}{\lambda} \,\frac{32\beta_\gamma^2\,{\rm ls}^{1/2} B^2}{\mpl^2\omega_{\rm pl}^3}\,\omega^{3/2}\,\mathcal{I}\left(\frac{\ell_\omega}{\lambda}\right)\,.
\end{equation}

The computation we have just outlined is valid as long as  the effect of the disformal coupling on the scalar field profile in the Sun is negligible. In other words, that the back-reaction effect due to the disformal coupling can be neglected. The disformal coupling leads to a contribution to the kinetic term of the scalar field proportional to $P/M_i^4$ \cite{Zumalacarregui:2012us,Sakstein:2014isa,Sakstein:2014aca,Ip:2015qsa,Sakstein:2015jca}, with $P \propto T^4$ the pressure of the solar photon gas, at a temperature $T$ (where, in our units, the proportionality factor is smaller than unity). Therefore, requiring that the disformal coupling does not modify the field-profile in the Sun is tantamount to requiring that $M_i \gtrsim T_{\rm core}$, and therefore $M_i \gtrsim {\cal O}({\rm keV})$. The condition in the tachocline is weaker as its temperature is lower than the core temperature by an order of magnitude.

In the  analysis of Sec.~\ref{sec:analysis}, we have imposed priors which ensure that $M_{\gamma}\,,M_e \gtrsim {\cal O}({\rm keV})$, to satisfy the previous bound. In addition, we note that for the benchmark point of parameter space we have discussed throughout the paper, and in particular in Figs.~\ref{fig:fig1} and~\ref{fig:fig2}, we have set $M_e = 10^{3.6}\,{\rm keV}$ and $M_{\gamma} = 1000\,{\rm TeV}$, such that the bound is  well satisfied.

\section{Chameleo-electric cross-section}
\label{sec:detectioncrosssection}

In this Appendix we derive an expression for the detection cross-section of chameleons from the analogue of the photoelectric effect. The cross-section for the ``chameleo-electric'' effect receives contributions from each of the three terms in the last line of the Lagrangian in Eq.~\eqref{eq:DEEFT2}. We first consider the disformal coupling between $\phi$ and the electron,
\begin{equation}
	\label{eq:disformal}
	\mathcal{L} \supset \sqrt{-g}\frac{1}{M_e^4}\partial_\mu \phi \partial_\nu \phi\,T_e^{\mu\nu}\,,
\end{equation}
where $M_e=M/d_e^{1/4}$ is the energy scale related to the disformal coupling with electrons. The stress-energy tensor associated with an electron four-spinor $\psi$ is
\begin{equation}
	T_e^{\mu\nu} = \frac{i}{2}\left(\bar\psi \gamma^{(\mu} D^{\nu)} \psi - D^{(\mu} \bar\psi \gamma^{\nu)} \psi \right)\,,
\end{equation}
where $\gamma^\mu$ are the Dirac matrices and a bracket denotes a symmetrisation over the four-indices $\mu$, $\nu$. In the following, we adopt Feynman slash notation $\not\! A =\gamma^\mu A_\mu$ for a four-vector $A_\mu$, and we define $\bar\psi = \psi^\dag\gamma^0$.

We decompose the electron free field as
\begin{equation}
	\psi = \sum_s \int \frac{\not\!\mathrm{d}^3{\bf p}}{\sqrt{2E_p}}\,\left(u_s({\bf p}) b_s e^{-ip \cdot x} + v_s({\bf p}) c_s e^{ip \cdot x}\right)\,,
\end{equation}
where the subscript $s$ labels the spinor component and $u_s({\bf p})$ and $v_s({\bf p})$ are Dirac spinors following the normalisation condition $\sum_s u_s\bar u_s = \not\!p +m_e$ and $\sum_s v_s\bar v_s = \not\!p - m_e$. The operator $b_s({\bf p})$ and its adjoint satisfy the anti-commutation relation
\begin{equation}
	\{b_s({\bf p}), b_{s'}^\dag({\bf p'})\} = \delta^{(3)}({\bf p}-{\bf p'})\delta_{ss'}\,,
\end{equation}
and similarly for the operator $c_s({\bf p})$, where curly brackets denote the anti-commutation of the two operators. We have defined $\not\!\mathrm{d}^3{\bf p} = \mathrm{d}^3{\bf p}/(2\pi)^3$.

The non-relativistic electron bound state is
\begin{equation}
	\psi = \sum_s \int \not\!\mathrm{d}^3{\bf p}\, \chi_s\varphi({\bf p}) b_s e^{-ip \cdot x}\,,
\end{equation}
where the Dirac spinor $\chi_s({\bf p})$ has the antiparticle entries equal to zero and $\varphi({\bf p})$ is a non-relativistic wave function in the momentum representation. We consider the ground state of a bound electron,
\begin{equation}
	\varphi(r) = \frac{1}{\sqrt{\pi}}\left(\frac{Z}{a_0}\right)^{3/2}\,e^{-Zr/a_0}\,,
\end{equation}
where $Z$ is the atomic number ($Z = 131$ for xenon) and $a_0$ is the Bohr radius. In momentum space, we obtain
\begin{equation}
	\varphi(p) \!=\! \int\!\! d^3{\bf r} e^{-i{\bf p}\cdot{\bf r}}\varphi(r) \!=\! \frac{8\sqrt{\pi}}{\left(p^2+(Z/a_0)^2\right)^2}\!\!\left(\!\frac{Z}{a_0}\right)^{5/2},
\end{equation}
where the wave function is normalised such that
\begin{equation}
	\label{eq:normmomentum}
	\int \not\!d^3{\bf p}\, |\varphi(p)|^2 = 1\,.
\end{equation}

\begin{figure}[!ht]
    \includegraphics[width=0.8\linewidth]{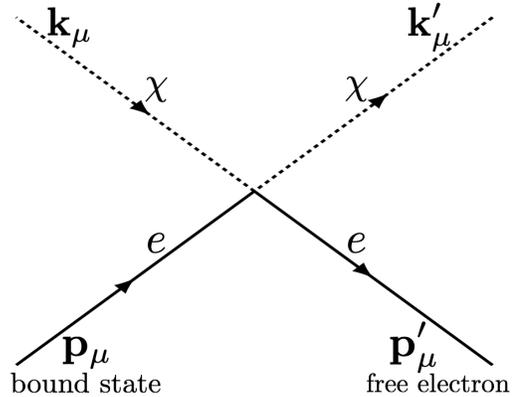}
    \caption{Feynman diagram for the scattering process associated with the disformal coupling in Eq.~\eqref{eq:disformal}.}
    \label{fig:fig6}
\end{figure}
\begin{figure}[!ht]
    \includegraphics[width=0.8\linewidth]{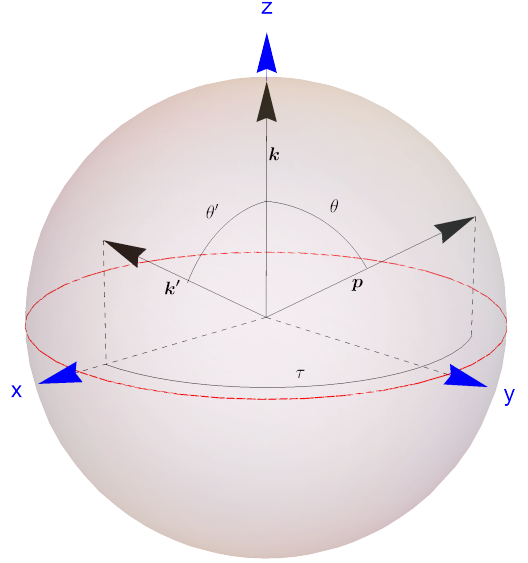}
    \caption{Relative orientation of the vectors ${\bf k}$, ${\bf k'}$, ${\bf p}$, in the spherical coordinate system chosen.}
    \label{fig:fig7}
\end{figure}

The scattering vertex from the disformal coupling in Eq.~\eqref{eq:disformal} is sketched in the Feynman diagram in Fig.~\ref{fig:fig6} and amplitude given by
\begin{equation}
	\mathcal M = -\frac{1}{4M_e^4}\,\bar u({\bf p'})\,\gamma^\mu\,{\rm A}_\mu\, \chi\,\varphi({\bf p})\,,
\end{equation}
where we have introduced the vector
\begin{equation}
	{\rm A}_\mu = k'_\mu\,(p+p')^\nu k_\nu + k_\mu\,(p+p')^\nu k_\nu'\,.
\end{equation}
The square of the amplitude summed over the spins of the final states and averaged over the spins of the initial states is
\begin{eqnarray}
	\label{eq:amplitude2}
	|\mathcal M|^2 &=& \frac{|\varphi(p)|^2}{64M_e^8}\,{\rm Tr}\left(\not\! {\rm A}\, (\not\! p+m_e)\not\! {\rm A} \,(1+\gamma^0)\right) = \nonumber \\
	&=& \frac{|\varphi(p)|^2}{16M_e^8}\left[ {\rm A}^2\, ( m_e \!-\! E') \!+\! 2{\rm A}^0\,({\rm A}^\mu p'_\mu)\right]\,,
\end{eqnarray}
with the cross-section
\begin{eqnarray}
	\label{eq:crosssect0}
	\sigma_{\phi e, {\rm dis}} &=& \int\! \frac{\not\!\mathrm{d}^3{\bf p}}{2k}\frac{\not\!\mathrm{d}^3{\bf k}'}{2\omega'}\frac{\not\!\mathrm{d}^3{\bf p}'}{2E'}|\mathcal M|^2\,\not\!\delta^{(4)}(k\!+\!p\!-\!k'\!-\!p') = \nonumber\\
	&=& \int\! \frac{\not\!\mathrm{d}^3{\bf p}}{2k}\frac{\not\!\mathrm{d}^3{\bf k}'}{2\omega'}\frac{1}{2E'}|\mathcal M|^2\not\!\delta(\omega\!+\!E\!-\!\omega'\!-\!E'\!) = \nonumber\\
	&=& \frac{1}{\left(2\pi\right)^5}\! \int\!\! \mathrm{d}y\mathrm{d}y'\mathrm{d}\tau \mathrm{d}p p^2\!\frac{(k')^2}{8kE'\omega' }|\mathcal M|^2\,\frac{\mathrm{d}k'}{\mathrm{d}\omega'}\,.
\end{eqnarray}
Here, $\not\!\delta(x) = 2\pi \delta(x)$ and we decomposed the four-vectors as follows. We consider the incoming and outgoing chameleon four-vectors $k^\mu =\left(\omega, {\bf k}\right)$ and $(k')^\mu=\left(\omega', {\bf k'}\right)$. The electron is described by the systems $p^\mu = \left(E, {\bf p}\right)$ and $(p')^\mu =\left(E', {\bf p'}\right)$, where $E = m_e - E_b$ and $E_b$ is the binding energy of the atomic electron. In the following, we make use of the magnitudes $k = |{\bf k}|$, $k' = |{\bf k'}|$, $p = |{\bf p}|$, and $p' = |{\bf p'}|$. We also introduce the angle $\theta$ between ${\bf k}$ and ${\bf p}$, the angle $\theta'$ between ${\bf k}$ and ${\bf k'}$, and the azimuthal angle $\tau$ between the projections of ${\bf k'}$ and ${\bf p}$ on the plane orthogonal to ${\bf k}$. The orientations of these vectors are sketched in Fig.~\ref{fig:fig7}.

Since $\mathrm{d}k'/\mathrm{d}\omega' = \omega'/k'$, the expression for the cross-section in Eq.~\eqref{eq:crosssect0} is
\begin{eqnarray}
	\label{eq:crosssection}
	\sigma_{\phi e, {\rm dis}} &=& \frac{1}{128\,M_e^8\,\left(2\pi\right)^5}\int \mathrm{d}y \mathrm{d}y' \mathrm{d}\tau\,\mathrm{d}p\,p^2|\psi(p)|^2\,\times \nonumber\\
	&& \frac{k'}{E'\,k}\left[A^2\,(m_e-E') + 2A_0 \,\left(A^\mu p'_\mu\right)\right]\,,
\end{eqnarray}
where we have used the expression for the amplitude squared in Eq.~\eqref{eq:amplitude2}.

To proceed with the computation, we consider the conservation of the four-vector on shell
\begin{eqnarray}
	E \!+\! \sqrt{k^2+m_\phi^2} \!&=&\! \sqrt{(p')^2\!+\!m_e^2} \!+\! \sqrt{(k')^2\!+\!m_\phi^2}\,,\label{eq:energycons}\\
	(p')^2 \!-\! (k')^2 \!-\! p^2 \!&=&\! k^2 \!-\! 2pk'x\!+\!2pky\!-\!2kk'y'\,,\label{eq:momentumcons}
\end{eqnarray}
where $y = \cos\theta$, $y' = \cos\theta'$, $x = yy' + \sin\theta\sin\theta'\cos\tau$. In the detector, the mass of the chameleon is expected to be set by a resonance condition involving the size of the cavity $R$~\cite{Khoury:2003rn,Brax:2007hi,Brax:2012gr}, and thus to be of the order of $m_\phi \sim 1/R \approx 10^{-7}\,{\rm eV}$, which is much smaller than other energies in the system. For this reason, we neglect $m_\phi$ in the rest of the computation. Combining Eqs.~\eqref{eq:energycons} and~\eqref{eq:momentumcons} we obtain
\begin{equation}
	|k'| = \frac{E^2 + 2 k E - p^2 - m_e^2 - 2 k p y}{2 \left(k + E - p x - k y'\right)} \approx k\,,
\end{equation}
where ``$\approx$'' indicates the limit $m_e \gg k \gg |E_b| \gg m_\phi$. We define the product
\begin{eqnarray}
	\alpha_1 &=& (p+p')^\nu k_\nu = (E+E')\omega - ({\bf p}+{\bf p'})\cdot {\bf k} = \nonumber \\
	&=& (E' \!+\! E)\omega \!-\! \left(2py \!+\! k \!-\! k'y'\right)k\,,
\end{eqnarray}
where in the last line the spatial part is ${\bf p'} = {\bf p} + {\bf k} - {\bf k'}$. Similarly, we define
\begin{eqnarray}
	\alpha_2 &=& (p+p')^\nu k_\nu' = (E+E')\omega' - ({\bf p}+{\bf p'})\cdot {\bf k'} = \nonumber \\
	&=& (E' \!+\! E)\omega' \!-\! \left(2px \!-\! k' \!+\! ky'\right)k'\,.
\end{eqnarray}
We then have $A_\mu = \alpha_1\,k'_\mu + \alpha_2\,k_\mu$. In the limit considered, $E \approx E' \approx m_e$, so we obtain
\begin{equation}
	\label{eq:approx}
	\alpha_1 \approx \alpha_2 \approx 2m_e\,\omega\,.
\end{equation}
In the definitions of $\alpha_1$ and $\alpha_2$, the temporal part of the four-product is the dominant one in the limit considered. The time component of the vector $A_\mu$ is
\begin{equation}
	\label{eq:A0}
	A_0 = \alpha_1\,\omega' + \alpha_2\,\omega\ \approx 4m_e\,\omega^2\,,
\end{equation}
where in the last step we used Eq.~\eqref{eq:approx}. The square of the vector $A_\mu$ is
\begin{eqnarray}
	\label{eq:A2}
	A^2 &=& A^\mu A_\mu = \left(\alpha_1 k'^\mu + \alpha_2 k^\mu\right)\left(\alpha_1 k'_\mu + \alpha_2 k_\mu\right) = \nonumber\\
	&=& m_\phi^2\left(\alpha_1^2 + \alpha_2^2\right) + 2 \alpha_1 \alpha_2\left(\omega\omega' - kk'y'\right) \approx \nonumber\\
	&\approx& 2 \alpha_1 \alpha_2\left(\omega\omega' - kk'y'\right)\,,
\end{eqnarray}
where the last approximation assumes a massless chameleon at detection, $k^\mu k_\mu = 0$ or $|{\bf k}| = \omega$. Since $\omega' \approx \omega$, in the limit considered we have
\begin{equation}
	\label{eq:A2p}
	A^2 \approx 8m_e^2\,\omega^4\,\left(1 - y'\right)\,.
\end{equation} We evaluate the product
\begin{eqnarray}
	\label{eq:Ap}
	A^\mu p'_\mu &=& A_0E' - {\bf A}\cdot {\bf p'} = \nonumber\\
	&=&\! (\alpha_1\,\omega' \!+\! \alpha_2\,\omega)E' \!-\! ( \alpha_1\,{\bf k'} \!+\! \alpha_2\,{\bf k})\cdot ({\bf p} + {\bf k} - {\bf k'}) = \nonumber \\
	&=&\! \left[\alpha_1\omega'(E' \!-\! px \!+\! ky' \!-\! k') \!+\! \alpha_2\omega(E \!-\! py \!+\! k \!-\! k'y')\right]\approx\nonumber\\
	&\approx& 4m_e^2\,\omega^2\,,
\end{eqnarray}
where in the last step $E \approx E' \approx m_e \gg |{\bf p}|, |{\bf k}|, |{\bf k'}|'$. Using the approximation
\begin{equation}
	E' \approx m_e + (1 - y')\,\frac{\omega^2}{m_e}\,,
\end{equation}
together with the expressions in Eqs.~\eqref{eq:A0},~\eqref{eq:A2p}, and~\eqref{eq:Ap}, we find
\begin{equation}
	A^2\,(m_e - E') + 2A_0 \,\left(A^\mu p'_\mu\right) \approx -8m_e\,\omega^6\,\left(1 \!-\! y'\right)^2 \!+ 32m_e^3\,\omega^4\,,
\end{equation}
so the term is dominated by the part $2A_0 \,\left(A^\mu p'_\mu\right)$. The second line in the computation of the cross-section in Eq.~\eqref{eq:crosssection} is then
\begin{equation}
	\label{eq:term2}
	\frac{k'}{E'\,k}\left[A^2\,(m_e-E') + 2A_0 \,\left(A^\mu p'_\mu\right)\right] \approx 32\,m_e^2\,\omega^4\,.
\end{equation}
For a chameleon produced in the Sun with an energy $\omega$, and in the limit in which its effective mass in the detector can be neglected, the cross-section is
\begin{eqnarray}
	\label{eq:crossection0}
	\sigma_{\phi e, {\rm dis}} &\approx& \frac{1}{128M_e^8\left(2\pi\right)^5}\int \mathrm{d}y \mathrm{d}y' \mathrm{d}\tau\,\mathrm{d}p\,p^2|\psi(p)|^2\,[32\,m_e^2\,\omega^4] = \nonumber\\
	&=& \frac{m_e^2\,\omega^4}{M_e^8\,\left(2\pi\right)^4}\int \mathrm{d}p\,p^2|\psi(p)|^2 = \frac{m_e^2\,\omega^4}{8\pi^2\,M_e^8}\,,
\end{eqnarray}
where the angular integral is trivial as there are no angles appearing in Eq.~\eqref{eq:term2}. In the last step, we have normalised the wave function according to Eq.~\eqref{eq:normmomentum}.

A second contribution to the cross section comes from the conformal term in Eq.~\eqref{eq:DEEFT2}
\begin{equation}
	\label{eq:trace}
	\mathcal{L} \supset \sqrt{-g}\beta_e\frac{\phi}{\mpl}T_e\,,
\end{equation}
for which the absorption cross section depends on the photo-electric cross section $\sigma_{\rm photo}$ in the limit $\omega \gg m_\phi$ as~\cite{Dimopoulos:1986mi, Dimopoulos:1986kc, Pospelov:2008jk}
\begin{equation}
	\label{eq:abs_crossection}
    \sigma_{\phi e, {\rm conf}} = \frac{\beta_e^2\omega^2}{2\pi\alpha \mpl^2}\,\sigma_{\rm photo}\,.
\end{equation}
We have taken the energy-dependent photoelectric cross section from Ref.~\cite{Veigele:1973tza}. Note, that we have not considered the production/detection from the XT coupling and therefore set $c_e = 0$.

In terms of the parameters used in the MCMC analysis, Eqs.~\eqref{eq:crossection0} and~\eqref{eq:abs_crossection} combine to give the following expression for the cross-section
\begin{equation}
	\label{eq:chameleoelectric}
	\sigma_{\phi e} = \sigma_{\phi e, {\rm dis}} + \sigma_{\phi e, {\rm conf}}= \frac{m_e^2\omega^4}{8\pi^2M_e^8} + \frac{\beta_e^2\omega^2}{2\pi\alpha \mpl^2}\,\sigma_{\rm photo} \,.
\end{equation}
Although in this work we have focused on xenon-based detectors such as XENON1T, the results of the computation can be applied more broadly to any material.

When the second term can be neglected, the event rate in the detector given by Eq.~\eqref{eq:diffrate_xenon}, with flux given in Eq.~\eqref{eq:production2}, gives
\begin{equation}
    \label{eq:diffrate_xenon0}
    \frac{\mathrm{d}R_0(\omega)}{\mathrm{d}\omega} = N_{\rm Xe}\,p_{\gamma, t}\,n_{\gamma, t}\,\frac{\Delta R}{\lambda} \frac{R_\odot^2}{d_\odot^2}\frac{\beta_{\rm eff}^2\,{\rm ls}^{1/2} B^2}{\pi^{3/2}\mpl^2\omega_{\rm pl}^3}\frac{m_e^2\omega^{11/2}}{{\rm keV}^8}\,,
\end{equation}
where $\beta_{\rm eff}$ has been defined in Eq.~\eqref{eq:betaeff}.

\bibliographystyle{apsrev4-1}
\bibliography{xenon}

\end{document}